%% file: REVISED_manuscript.tex
\documentclass[12pt]{iopart}
\usepackage[utf8]{inputenc}
\usepackage{booktabs}
\usepackage[T1]{fontenc}
\usepackage{graphicx}
\usepackage{hyperref}
\usepackage{soul,color}
\usepackage{geometry}
\usepackage{tabularx} 
\usepackage{pdflscape}
\usepackage{natbib}
\usepackage{longtable}
\usepackage{booktabs}
\usepackage{adjustbox}
\usepackage[table,xcdraw]{xcolor}
\usepackage{url}
\usepackage{comment}

\usepackage{ragged2e}

\bibpunct[, ]{(}{)}{,}{a}{}{,}%
\usepackage{etoolbox}
\makeatletter
\newcommand{\mainmatter}{%
  \setcounter{footnote}{0}%
  \patchcmd{\@makefntext}{\fnsymbol}{\arabic}{}{}%
  \patchcmd{\@thefnmark}{\fnsymbol}{\arabic}{}{}%
  \def\@makefnmark{\textsuperscript{\arabic{footnote}}}%
}
\makeatother

\begin{document}

\title[Economic complexity and the sustainability transition]{Economic complexity and the sustainability transition: A review of data, methods, and literature}

\author{Bernardo Caldarola\textsuperscript{\textdagger}, Dario Mazzilli, Lorenzo Napolitano\textsuperscript{\textdaggerdbl}, Aurelio Patelli, Angelica Sbardella}

\address{Enrico Fermi Research Center, via Panisperna 89a, 00184, Rome (Italy)}
\address{\textsuperscript{\textdagger}UNU-MERIT and University of Maastricht (Netherlands)}
\address{\textsuperscript{\textdaggerdbl}European Commission Joint Research Centre (JRC-Seville)}
\ead{aurelio.patelli@cref.it}
\begin{abstract}
Economic Complexity (EC) methods have gained increasing popularity across fields and disciplines. In particular, the EC toolbox has proved particularly promising in the study of complex and interrelated phenomena, such as the transition towards a greener economy. Using the EC approach, scholars have been investigating the relationship between EC and sustainability, proposing to identify the distinguishing characteristics of green products and to assess the readiness of productive and technological structures for the sustainability transition. This article proposes to review and summarize the data, methods, and empirical literature that are relevant to the study of the sustainability transition from an EC perspective. 
We review three distinct but connected blocks of literature on EC and environmental sustainability. First, we survey the evidence linking measures of EC to indicators related to environmental sustainability. Second, we review articles that strive to assess the green competitiveness of productive systems. Third, we examine evidence on green technological development and its connection to non-green knowledge bases. Finally, we summarize the findings for each block and identify avenues for further research in this recent and growing body of empirical literature.  
\end{abstract}

\newpage
\mainmatter

\section{Introduction}\label{intro}

The notion of Economic Complexity (EC) has been widely used to encompass a set of methods that characterize the productive and technological composition of economies (countries, regions, cities) relying on complex systems science \citep{Hidalgo2009, Tacchella2012}. EC methods have proved to be particularly effective in predicting future patterns of economic growth using information on the export basket of countries \citep{Tacchella2018}. The key intuition behind the EC approach is that economic development and growth are the result of the specialization and diversification patterns of economies, which emerge from underlying hidden interactions between elements in the society \citep{Balland2022,pugliese2019unfolding}. More specifically, EC focuses on the role played by the accumulation of (unobserved) productive and technological capabilities in driving economic diversification and growth \citep{hausmann2007you,Hausmann2011atlas,pugliese2017complex,sbardella2018role}. By preserving information on \textit{what} economies produce, rather than merely \textit{how much}, the EC literature is able to ``describe and compare economies in a manner that eschews aggregation'' \citep{hausmann2011network}. The EC approach yields a complementary perspective on several domains of economically relevant human activity (e.g. trade, technical innovation, scientific research) with respect to conventional (i.e. aggregate) indicators on productive inputs or performance. This emphasis on the content of the activity baskets of countries or regions resonates with other approaches in economics that understand economic growth through the lens of sectoral allocation of productive factors, such as the structuralist literature \citep{Prebisch1950, Lewis1954, Hirschman1958, Lin2011}, and that identify in the role of capabilities the main factor driving innovation, embodied in the evolutionary economics literature \citep{Cimoli1995, Dosi1994, Nelson1982, Teece1994}. 

The quality and diversity of an economy's productive and technological portfolio have broader implications than simply economic growth. For instance, the composition of economic and technological specialization has strong implications for the environment, as the footprint of different products and technologies can differ substantially. By the same token, the accumulation of technological capabilities may put countries on trajectories that can mitigate or exacerbate climate change. In this respect, EC methods can prove particularly useful to understand and guide the sustainability transition. 

The transition towards a more sustainable and low-carbon socioeconomic system is a top policy priority, with the EU Green Deal aiming at climate neutrality by 2050\footnote{\url{https://eur-lex.europa.eu/legal-content/EN/TXT/?qid=1596443911913\&uri=CELEX\%3A52019DC0640\#document2}}, the Inflation Reduction Act in the United States \citep{bistline2023economic} and the ambitious renewable energy targets of China's 14th Five-Year Plan For Renewable Energy With Quantitative Targets For 2025 \citep{li2019economic,yang2023potential}. However, decarbonizing the economy by phasing out high-emission and energy-intensive industries will require radical transformations and profound structural change at the core of socioeconomic systems. Moreover, this will have to account for the heterogeneous capacity of geographical areas and industries to achieve climate neutrality along with the possible long-lasting effects on income, spatial, and environmental inequalities.

To inform policy on how to address such a complex transformation, wherein geographical, structural and institutional elements interact, in the last few years an approach has emerged in the literature (see among others \citealp{barbieri2020specialization,barbieri2022regional,montresor2020green,santoalha2021diversifying}) that draws from sustainability studies, evolutionary economic geography \citep{boschma2006economic,boschma2018evolutionary} and EC. Embracing a complexity perspective may in fact be more effective than traditional approaches in accounting for the interconnected nature of this process of change \citep{common2005ecological} providing policy-relevant, data-driven and granular evidence to embrace socioeconomic complexity at different geographical scales. 

While environmental goods and technologies have been mainly studied as homogeneous, aggregate quantities in the extant literature,  they actually display high degrees of heterogeneity in terms of functions, underlying know-how, life cycle stages, and links to pre-existing specialisation patterns. Therefore, the capacity of EC methodologies to keep away from aggregation and to provide feasibility diagnostics at the level of single products or technologies 
may prove extremely relevant in analyzing the potential directions of green development for each country or region,  

Coupling the geographical distribution of productive and technological capabilities with environmental and socioeconomic variables, various scholars have applied the EC toolbox to try to answer three main broad questions:
(i)	What is the relationship between complexity and sustainability? (ii) What are the properties of green products and technologies, and are these inherently different from non-green ones? (iii) How can we assess the readiness of national/regional knowledge bases and productive structures for the green transition? 

Contributions addressing the first question investigate the empirical relationship between countries’ environmental outputs, such as Greenhouse Gases (GHG) emissions, and the complexity of their productive structures, assessed by focusing on the export dimension through various EC metrics. These studies are far from unanimous, as they vary in their samples or estimating methods. This yields contrasting results on the effect of EC on the environment, ranging from linear positive to non-linear relationships \textit{à la} Kuznets curve \citeyearpar{Kuznets1955}.

The second and third questions have roots in the evolutionary economic geography literature and are more varied in terms of techniques and scales of observation. They mainly examine the geography of green competitiveness of national/regional knowledge bases or productive structures by analyzing specialization profiles in green products or technologies, different dimensions of relatedness between green and non-green technological or productive activities, and their relationship with a series of socioeconomic indicators -- e.g., inequality or digital literacy. As is common in the EC literature, these works especially stress the potential complementarity between existing capabilities and future specialization in low-carbon technologies or products. 

The EC approach emphasises the role played by productive and technological capabilities in promoting advancement across domains of human activity. 
From the point of view of green goods and innovative activities, this implies that as countries and regions strive to move towards sustainable practices, EC offers a framework to analyze the composition and diversification of their economies, and to identify the dynamic set of capabilities that are most conducive to green productive or technological specialization, allowing countries to progress along the sustainability ladder. As we show in this review, EC provides a valuable lens through which policymakers and researchers can understand and navigate the intricacies of the sustainability transition.

This literature review aims to summarize the empirical evidence produced so far to answer the three aforementioned questions from an EC perspective. It should be noted that other recent contributions have tackled the link connecting EC and socioeconomic sustainability from several angles. Notably, the recent review by \cite{montiel2024intrinsic} focuses on the ways in which EC has proven to be useful to relevant aspects of the economic literature such as economic growth, innovation, employment, education levels, poverty reduction, life expectancy, environmental degradation, public policies, etcetera. The paper identifies five areas of interest for EC, namely investigating the causal pathways linking EC to sustainability dimensions; developing new indicators of EC; exploring how EC can be used to promote sustainable development policies as well as international research cooperation on this topic; and integrating EC into tertiary education programs. At the same time, several contributions have ventured into a detailed analysis of how EC indicators correlate with empirical measures of socioeconomic sustainability (some of which are reviewed by \citet{Hidalgo2021}, and more extensively in Section \ref{eci-environ} of this paper). What makes this review complementary to the aforementioned contributions is that it focuses explicitly on the methodologies and data that have been used to link EC to environmental sustainability only, in an attempt to answer the macro-questions listed above. Our effort is particularly timely given the rapid expansion of this literature, and it provides scholars with a clear roadmap on how to produce results that are comparable to the ones produced in the recent past.

The remainder of this paper is organized as follows. We first review the data on products and technologies used to apply EC methods to the study of the sustainability transition, highlighting drawbacks and advantages entailed by different types of data (Section \ref{data}). Second, we offer a methodological contribution by attempting to unify the methods used to estimate and analyze measures of EC and relatedness (Section \ref{methods}). Third, we canvas the literature that links complexity measures to the concept of environmental sustainability, in order to summarize the debate on the role of complexity in explaining environmental issues and in identifying viable avenues into the sustainability transition (Section \ref{review}). Finally, we attempt to identify limitations of the existing literature and methods, proposing further research avenues in the field (Section \ref{conclusions}).
   
\section{Data and Methods}\label{data-methods}

The data employed in EC-based analysis of the green economy are drawn from two main sources: patents and trade flows. This section presents in detail the most commonly used datasets and classifications of green activities, firstly focusing on patent data and secondly on trade data. This data is used in EC following a common empirical framework: a bipartite matrix is constructed by cross-tabulating products (technologies) and geographical areas (countries, regions, cities...). The bipartite matrix is then filtered, projected, ordered, and processed in different ways, depending on the specific purposes of the analysis (see Section \ref{methods}).

\subsection{Data}\label{data}

\subsubsection{Patent data}\label{patent-data}

\input{section-patent_data}

\subsubsection{Trade data}\label{trade-data}
Since \textit{The environmental goods and services industry manual for data collection and analysis} was published by the \citet{OECD1999}, a wide array of lists and taxonomies of green products has been proposed. Here we briefly discuss the lists that are most commonly used in EC analyses, highlighting the main critical issues in classifying environmental goods. Readers should keep in mind that the EC literature has focused, especially in earlier works, on establishing the level of complexity and growth potential of an economy by extracting information from its trade specialization profile. In order to do so, a harmonized global classification of products with a uniform, global interpretation has been crucial to develop the framework. The Harmonized System (HS) classification satisfies these requirements and has been the most widely used in the field. HS is a standardized numerical method of classifying traded products used by customs authorities to identify products; it is maintained by the World Customs Organization (WCO) and it is updated every five years. HS comprises more than 5,000 commodity subheadings, which are identified by a 6-digit code and arranged according to a nested structure going up to 96 2-digit Chapters and 21 1-digit Sections. 

The first issue to take into account when constructing a comprehensive classification of green products is in the very nature of the HS classification system, which was not conceptualized to accommodate a green/non-green dichotomy or information on the goods' efficiency. 
The introduction in the HS of several 6-digit subheadings which group together new environmental goods was announced \citep{steenblik2020code} in 2020; the WCO is slowly addressing this issue by updating the HS system with more codes for green products. However, the updated classification is not yet available. Therefore, a clear-cut identification of environmental goods within existing product classifications may at times prove to be a difficult task. For instance, up until recently, it was impossible to distinguish between combustion engines and electric vehicles.
Other classifications or surveys exist, although they focus on specific environmental aspects or on final use, and do not present the standardization and granularity properties required by EC techniques.  A second crucial requirement for a comprehensive and effective green goods classification is a unanimous definition of what \textsc{green products} are. This clearly depends on specific needs: for instance, different requirements need to be satisfied to fulfil regulatory and research goals, or to define (tax) incentives towards more sustainable practices and energy sources. 

While stressing a series of open questions, partially still unanswered, a 1999 OECD report  \citep{OECD1999} defines the environmental goods and services industry as the set of activities aimed at measuring, preventing and mitigating environmental damages related to eco-systems, water, air, soil, and noise pollution, or waste management. This includes cleaner technologies, products and services that reduce environmental risk and minimize pollution and resource use. In the report, a list of 121 environmental goods satisfying this definition was proposed. More recently, according to the International Monetary Fund (IMF)\footnote{The definition of environmental goods of the IMF Climate Change Indicator Dashboard, 'Trade in Environmental Goods' data-set can be found at: \url{https://climatedata.imf.org/datasets/8636ce866c8a404b8d9baeaffa2c6cb3_0/explore}, where international trade flows in environmental goods -- defined according to the OECD 1999's list \citep{OECD1999} as updated by the IMF -- are reported.}  environmental goods are connected to environmental protection or goods that have been adapted to be more environmentally friendly. 
To achieve an encompassing classification for policy implications, evaluation, and tariff regularization, in 2009 the WTO published a broader list of green products \citep{WTO2009} comprising 480 products. While the list is comprehensive, many inaccuracies and biases have been pointed out by a EUROSTAT report published in the same year \citep{Eurostat2009}. 
The WTO also published two shorter \textit{friend} and \textit{core} lists composed respectively of 154 and 26 products. Similar efforts were made by the OECD for the 2010 Toronto G20 summit which proposed an updated list of 150 Plurilateral Environmental Goods and Services (PEGS). Moreover, in 2012 the Asian-Pacific Economic Cooperation (APEC)  \citep{APEC2012} put forward a list of 54 environmental goods subject to reduced tariffs. Finally, by combining and revising the WTO friend, APEC, and PEGS lists, in 2014 the OECD published the most recent list of green products, the Combined List of Environmental Goods (CLEG) \citep{CLEGS2014} comprising 248 products relevant to tackling climate change. 

In addition to the fact that any list of environmental products cannot be considered final as the HS classification needs constant updates and revision, regardless of the specific definition, two main shortcomings are shared by all green product classifications:
\begin{itemize}
    \item Final use:  it is not possible to ascertain the actual final use of a large number of products. Many commodities labelled as green (e.g., filters, pumps, and pipes) may be used also for non-environmental purposes. While statistics can be computed based on surveys, their reliability on a global scale is yet unclear.
    \item Greener products: depending on the goal of the study/application, one could consider products that are less raw material-intensive, more energy efficient, easier to dispose of, or have longer life spans. However, these properties crucially rely on the comparison with other products belonging to the same category, comparisons that are, even in principle, very difficult to carry out. 
\end{itemize}
Therefore, a clear trade-off between accuracy and comprehensiveness arises, and, as we will see below, the selected contributions using green products have proposed different identification strategies to, albeit not fully, overcome these issues.

\subsection{Methods}\label{methods}
The basic intuition of the EC approach is that specific activities, such as exported products, industries, patents, or scientific research are important because they constitute different learning opportunities and development possibilities. EC explicitly builds on the heterogeneity and interactions between different economic actors, assuming that 
knowledge grows not by accumulating ‘more’ of the same, but by adding new and different elements to existing capacities. In this framework, economic performance is seen as the result of the accumulation of different types of non-tradeable inputs and capabilities, that are not empirically observable but can be inferred by studying the structure of the specialization profiles of countries. Therefore, EC can be seen as an indirect measure of the capability endowment of a country. The notion of capabilities was developed at the firm level in the evolutionary economics literature to describe the dynamic know-how allowing firms to develop and introduce new products and services in the market \citep{penrose1959theory, teece1997dynamic,teece1994coherence}. In the EC literature, productive capabilities describe location-based attributes, which encompass further intangible aspects contributing to building and effectively exploiting productive efficiency, such as the institutional setups, education systems, policies, and infrastructures needed by a country to learn how to specialise in more complex activities \citep{hausmann2003economic, hausmann2005self,sutton2016capabilities}. Information on capabilities is gathered from a binary network that connects geographical areas to the activities in which they hold a comparative advantage.

In the following, we present the main metrics and tools used in the EC literature analyzing different dimensions of sustainability.

\subsubsection{Complexity measures}\label{sec:compl_measures}
EC analyses often start from the observation of empirical bipartite networks connecting geographical areas -- be they countries, regions, or cities -- to different types of economically relevant activities -- such as patenting \citep{balland2017geography,pugliese2019unfolding,sbardella2018green}, manufacturing of products \citep{Hidalgo2009,Tacchella2012}, or scientific research~\citep{patelli2023geography} -- in which they specialize.
These bipartite networks are typically assessed by evaluating the comparative advantage of the geographical area in the selected activity, using Balassa's Revealed Comparative Advantage (RCA) index~\citep{balassa1965trade}. RCA measures the share of an activity in a geographical area with respect to a reference distribution, often the activity's global share, and is thus interpreted as a proxy for above-average competitiveness in that activity. 
The index (and its refinements -- see~\citealp{Bruno2023}) remains the workhorse for binarizing the adjacency matrices of bipartite networks in the EC literature. In formula, the RCA of  geographic area $g$ in activity $a$ of volume $W$, can be written as follows:
\begin{equation}
    [RCA]_{ga} = \frac{W_{ga}(\sum_{g'a'}W_{g'a'})}{(\sum_{a'}W_{ga'})(\sum_{g'}W_{g'a})} = \frac{W_{ga}}{\sum_{a'}W_{ga'}} {\Large /} \frac{\sum_{g'}W_{g'a}}{\sum_{g'a'}W_{g'a'}}.
\end{equation}
RCA assigns a real and positive value to each combination of geographical areas and activities. $[RCA]_{ga}=1$ if the weight $W$ of $a$ in the basket of activities of geographical area $g$ is the same as the global weight of activity $a$ relative to all activities. 
The resulting distribution of RCA values is skewed with large tails \citep{patelli2022evolution}. 
High-variability continuous values can be impractical in most applications, especially when the only information required is on the presence/absence of specialization of geographical area $g$ into activity $a$. Therefore, the RCA matrix is usually binarized by setting a threshold value of 1.\footnote{Even though the intuitive unit threshold is standard in the EC literature, different values may be appropriate depending on the specific application.} A matrix entry equal to 1 is inserted in the binarized matrix cell when $[RCA]_{ga} \geq 1$ -- i.e., $g$ has a comparative advantage in $a$ --, while an entry equal to zero is inserted otherwise:

\begin{equation}
M_{ga}=\left\{
\begin{array}{l}
    1 \qquad if \, \, [RCA]_{ga} \geq 1 \\
    0 \qquad otherwise.
\end{array}
    \right.
\end{equation}

The binary RCA network displays strong empirical evidence of diversification and nestedness patterns in all the activities studied in EC: many geographical areas are competitive in large set of activities, in contrast to Ricardian specialisation patterns. However, there is more to the binary $M$ matrices than just the diversification of the geographical areas represented by their rows. $M$ matrices are in fact usually nested across domains of activity, irrespective of whether the columns of $M$ represent products, technologies, or scientific fields ~\citep{patelli2022evolution} and scales of analysis~\citep{laudati2023different,pugliese2019emergence,sbardella2017economic}. 
The concept of nestedness originates from ecology~\citep{atmar1993measure} and refers to the observation that less diversified geographical areas (species) typically have comparative advantages in the activities (ecological niches) pursued by more diversified geographical areas. This pattern can be made explicit when the presence/absence $M$ matrix is shown with the proper ordering of rows and columns, resulting in the emergence of a hierarchical model represented by triangular-like matrices~\citep{Cristelli2015}.
 
Based on the information contained in the binary matrix $M$, it is possible to define synthetic EC indicators capturing relevant information on the network structure, and thus on the geographical areas' specialization profiles. The rationale of EC algorithms is that the complexity of the economies under analysis and the complexity of the activities in which they are specialized can be determined recursively by taking advantage of the information provided by the composition of their activity portfolio. 
The first indicator of Economic Complexity was proposed by~\cite{Hidalgo2009}, the 'Method of Reflections' (MR) aimed at capturing the diversification of a country's export profile. At the $n$-th iteration, for a geographical area $g$ and activity $a$, MR proposes an iterative procedure to capture the diversification of geographical area $g$, $k_g^n$, and of activity $a$, $k_a^n$:
\begin{equation}
    \begin{array}{r@{}l@{\qquad}l}
	k_g^{(n)}=\frac{1}{k_g^0}\sum_a M_{ga} k_a^{(n-1)} & \\ \\
	k_a^{(n)}=\frac{1}{k_a^0}\sum_g M_{ga} k_g^{(n-1)} & \\
    \end{array}
    \label{eq:reflections}
\end{equation}
with initial conditions $k_g^0=\sum_a M_{ga}$ and $k_a^0=\sum_gM_{ga}$,  the diversification of $g$ and the ubiquity of $a$.
The MR rationale is that the complexity of a geographic area, interpreted as its generalized diversification, is driven by the average complexity of the activities it specializes in. By the same token, the average complexity of each activity is driven by the average complexity of the areas displaying a comparative advantage in it.
Since the iterative model in Equation~\ref{eq:reflections} has a trivial solution where each component is $1$~\citep{Caldarelli2012,Kemp2014}, the same authors later proposed a more practical definition of the indices based on the eigenvector associated with the second largest eigenvalue of a matrix derived from the MR \citep{mealy2019interpreting}:  the Economic Complexity Index (ECI) of countries, and the Product Complexity Index (PCI), in the case of products~\citep{Hausmann2011atlas,Hausmann2014atlas}.
Both indices converge to those evaluated using MR only after considering the standardized version (removing the mean and dividing by the standard deviation), although there is a certain carelessness in the economic complexity literature that refers to the ECI/PCI only as the second-largest eigenvalue.
Their generalization to the case of other activities, such as patents, is straightforward: the same mathematical procedure is applied to the binary matrix based on the data on the activity of interest. 

However, the ECI-PCI approach presents a non-negligible conceptual drawback due to the fact that ECI and PCI are defined by averages~\citep{Pietronero2017caciara} and not by sums, hence smoothing out the information on the cumulative nature of capabilities encapsulated in the geographical area-activity networks, a pillar of the economic reasoning behind the EC approach~\citep{Dosi1994,Teece1994}. This means that, for example, if we consider two fictitious countries, with the first specialized in all the activities present in the network, and the second holding a comparative advantage only in one activity of average complexity, they will paradoxically display the same ECI value.
Interestingly, \cite{mealy2018new} define a Green Complexity Index (GCI) simply as the sum of the complexity of green products (while product complexity is still calculated on the entire set of products using ECI), with the final sum accounting for country diversification in green products, partially overcoming this crucial aspect of ECI. 

While ECI is based on the idea that countries are diversified, the observation of nested patterns uncovers complex and interdependent systems that cannot be easily described through linear models and averages.
A more suitable metric for the characterization of nested patterns in the $M$ matrix is the Economic Fitness and Complexity (EFC) algorithm~\citep{Tacchella2012}: the Fitness of a geographical area $g$ is defined as the sum of the complexity of all the activities in which the area is specialized (i.e., if $RCA_{ga}\geq 1$). Therefore, Fitness increases as the activity basket of $g$ become larger: a geographical area with a more advanced set of capabilities will have a more diversified portfolio of activities, spanning from the most to the least complex ones, and will therefore have a higher Fitness score.\footnote{It has been shown that Fitness and Complexity can be interpreted as \textit{potential} functions, related to the bipartite network $M_{ga}$, which defines a forbidden region of the matrix given by the requirement of efficiency in the allocation of resources of countries and products~\citep{mazzilli2024equivalence}.} The extensive nature of Fitness is complemented by the definition of the Complexity of activity $a$, which is driven mainly by the lowest Fitness areas holding a comparative advantage in $a$, since complex activities are rare and appear almost exclusively in the portfolio of high-Fitness areas. Consequently, an area with low Fitness has a smaller endowment of capabilities and thus operates exclusively in less complex domains. Operationally, the EFC iterative algorithm is defined as the fixed point of the following coupled equations, that are suitably normalized at each iterative step:\footnote{The normalization step is unnecessary for the evaluation of the fixed point \citep{mazzilli2024equivalence}, however, it is helpful for the numerical procedures and the stability of the code.}
\begin{equation}
    \left\{\begin{array}{r@{}l@{\qquad}l}
	\widetilde{F}_g^{(n)}=\sum_a M_{ga} Q_a^{(n-1)} &\qquad & F_g^{(n)}=\frac{\widetilde{F}_g^{(n)}}{\langle\widetilde{F}_g^{(n)}\rangle_g} \\ \\
	\widetilde{Q}_a^{(n)}=\frac{1}{\sum_{g} M_{ga} \frac{1}{F_{g}^{(n)}}}  &\qquad & Q_a^{(n)}=\frac{\widetilde{Q}_a^{(n)}}{\langle\widetilde{Q}_a^{(n)}\rangle_a}
    \end{array}\right.
    \label{eq:f-q}
\end{equation}

It is noteworthy to mention that both Fitness and ECI are relative measures, based on the structure of the geographical area-activity matrix $M$ at a given time of observation of the economic system under analysis. Therefore, their magnitudes cannot be compared longitudinally across time as, at each given time point, the underlying bipartite networks change. This issue is not often discussed in the literature, but it should be a pressing concern when these metrics are included in longitudinal regression settings (many such examples are discussed in Section \ref{eci-environ}). It is however possible to overcome this issue by using ECI/Fitness rankings, albeit losing information on the EC index's magnitude, or more soundly by setting a reference scale~\citep{mazzilli2024equivalence}, i.e. by assigning to a reference country a fixed value, or by adding a dummy country that is always specialized in all activities, therefore imposing an invariant of scale across time periods.

Another frequent issue that appears in the EC literature is the need to analyze systems across different geographical scales. Often, at finer geographical scales (i) local data-set do not have global coverage; (ii) data quality is lower since volumes are smaller and errors become more relevant. 
While issue (ii) is straightforward, issue (i) is often overlooked, although it may pose problems in evaluating a reference distribution for computing the RCA (e.g., it is very different to assess the complexity of products when looking at global trade flows or only at European countries or regions). To overcome these issues, a measure of \textit{exogenous fitness} has been proposed~\citep{Operti2018,sbardella2018green}. To compute the Fitness of subnational areas, this approach computes the Complexity of economic activities at the country level, and subsequently evaluates the Fitness of subnational areas by plugging the country-level complexity in the computation of the Fitness equation (first term of Equation~\ref{eq:f-q}). In such a way, (i) the signal-to-noise ratio in the data is higher; (ii) it is possible to provide a more realistic assessment of complexity evaluating competition dynamics at the global level. In principle, the same rationale may be implemented in the ECI definition (Equation~\ref{eq:reflections}). 
In particular, Green Technology Fitness, introduced by \cite{sbardella2018green} and used in other works reviewed in the present paper, is a measure of Fitness computed on climate change mitigation technologies (see Section~\ref{data}), which, when computed at the regional level, employs the exogenous fitness rationale. 

\subsubsection{Relatedness}\label{sec:relatedness}
Another key set of techniques from the EC toolbox aims at measuring relatedness\footnote{These analytical tools can be traced back to \cite{jaffe1986technological} and to the measure of corporate coherence introduced by \cite{Teece1994}.}, a measure of the pairwise similarity between activities --  be they products \citep{Hidalgo2009ps,zaccaria2014taxonomy}\footnote{See also \cite{tacchella2023relatedness} for a multi-product, non-linear approach based on machine learning.}, technologies \citep{breschi2003knowledge,napolitano2018technology}, or both technologies and products \citep{pugliese2019unfolding,de2022trickle}. 
The measure of relatedness between two activities is based on the observation of their empirical co-occurrences in the specialisation profiles of different geographical areas, and is connected to the probability that having a comparative advantage in the first activity will also lead to a comparative advantage in the second. This is done relying on the assumption that countries that can successfully specialise in an activity have developed a set of capabilities that will enable them to diversify into related activities. Intuitively, activities that share similar inputs will be situated close to each other in the network, and proximity in the network should indicate a relatively high probability of jumping from an activity $a$ to a neighboring one, $a'$. 

Relatedness is usually represented as a network of connected activities, where links between activities are based on their degree of similarity. There are several methods to evaluate relatedness, resulting in different and heterogeneous networks that cannot always be easily compared. However, most of these networks present a core-periphery structure, with hubs and leaves associated with similar types of activities or arranged into meaningful communities.
The first approach to relatedness in the EC literature is the Product Space, proposed by~\citet{Hidalgo2007}, initially based on the country-product international trade network, and later employed also for different set of activities. The similarity between activity $a$ and activity $a'$ is evaluated using \emph{proximity}\citep{Hidalgo2009ps}, a measure of pairwise normalized geographical co-occurrence, i.e. summing over the geographical dimension $g$:
\begin{equation}\label{eq_proximity}
    \phi_{aa'}=\min\left\{  \frac{\sum_g M_{ga}M_{g,a'}}{\sum_c M_{ga}}, \frac{\sum_c M_{ga}M_{ga'}}{\sum_c M_{ga'}}\right\}.
\end{equation}
Originally, proximity was interpreted as the conditional probability of being competitive in activity $a$ given competitiveness in activity $a'$~(see p. 3 of \citealp{Hidalgo2007} SI).~\footnote{Please notice that even though each value of $\phi_{pp'}$ is real, positive, and bounded to 1, proximity is not a conditional probability. In fact, a conditional probability $P(A\vert B)$ requires that $\sum_AP(A\vert B)=1$, while $\sum_{p'} \phi_{pp'}$ depends on the dimension of the network (the linear sizes of $M_{cp}$) and is typically larger than 1.}

The EC literature has later proposed a more statistically sound approach to relatedness that allows to evaluate time-lagged connections between pairs of activities, also between two layers of different types of activities such as products and technologies, or technologies and scientific fields \citep{de2022trickle,pugliese2019unfolding,sbardella2022regional,zaccaria2014taxonomy}, allowing to model time-lagged spillovers across different dimensions of capability.  
Initially focused on international trade, \cite{zaccaria2014taxonomy} put forward a refined methodology to assess relatedness, the Product Progression Network. This approach has three main advantages: i) it is dynamic, as it takes explicitly into account time by comparing the pairwise activity co-occurrences within the same geographical area at different points in time, instead of contemporaneous co-occurrences; ii) it filters each link of the resulting network using a null model so that only the statistically significant links are stored: a crucial advancement, because the presence of a link may be simply due to the ubiquity of a product or to the diversification of a country~\citep{Cimini2019,Saracco2015,Saracco2017}; iii) instead of focusing on pairwise correlation between two activities of the same type (e.g. two products), it has been employed to study also the co-occurrence among different types of activities, such as links between products and technologies \citep{de2022trickle,pugliese2019unfolding}.

The mathematical formulation of this approach relies on the so-called \textit{assist matrix} $B$, which estimates the probability of having a comparative advantage in activity $a$ in geographical area $g$ in year $y_2$, conditional on having a comparative advantage in activity $a'$ in the same area in a previous year $y_1= y_2-\Delta t, \Delta t \geq 0$ (typically a lag of $\Delta t =5$ years is considered) and is defined as follows:
\begin{equation}
	B^{\alpha\beta}_{aa'}(y_1,y_2) = \sum_{g}\frac{M^{\alpha}_{ga}(y_1)}{u^\alpha_a(y_1)}\frac{M^\beta_{ga'}(y_2)}{d^\beta_g(y_2)}, \quad u^\alpha_a=\sum_g M^\alpha_{ga},\,\, d^\alpha_g=\sum_a M^\alpha_{ga}
 \label{eq:assit_matrix}
\end{equation}
where $\alpha$ and $\beta$ indicate that the matrices may contain information on different activity types (e.g., $\alpha$ being patents, and $\beta$ products).
In this case $B_{aa'}$ is defined as a conditional probability, since $\sum_{a'} B_{aa'}=1$, $0\le B_{aa'} \le 1$. 

However, the assist matrix, like any other relatedness measure, is based on nested networks, where diversified actors are active in most of the activities, and as a consequence, a co-occurrence may not be informative \textit{per se}.
Hence, a null-model should be employed to discount the fact that co-occurrences can be random, and are more likely for more ubiquitous activities or more diversified geographical areas \citep{Cimini2019}. The most common null model used in the literature to assess the statistical significance of the conditional probabilities $B_{aa'} (y_1, y_2)$, is the Bipartite Configuration Model (BiCM) \citep{Saracco2015,Saracco2017}, a maximum-entropy approach for the randomization of bipartite networks -- we refer to \citep{Cimini2019} for a more detailed discussion on null models for complex networks. 


\section{Review of the literature}\label{review}
In this section, we propose a selection of papers analyzing different dimensions of the sustainability transition using EC techniques. This exercise is focused not only on highlighting the main findings but especially on illustrating the variety of methods employed, which lead often to inconclusive results due to their low comparability.
For this reason, we conduct a purposive and ad hoc literature review, with the aim of illustrating the empirical applications of the methods discussed in the previous section. Systematically reviewing all the evidence on the use of EC metrics and methods to study the sustainability transition falls beyond the scope of this article; instead, we strive to offer a representative account of the empirical EC applications in the field. 

\begin{landscape}
\begin{figure}
    
    \centering
    \includegraphics[width=1.25\textwidth]{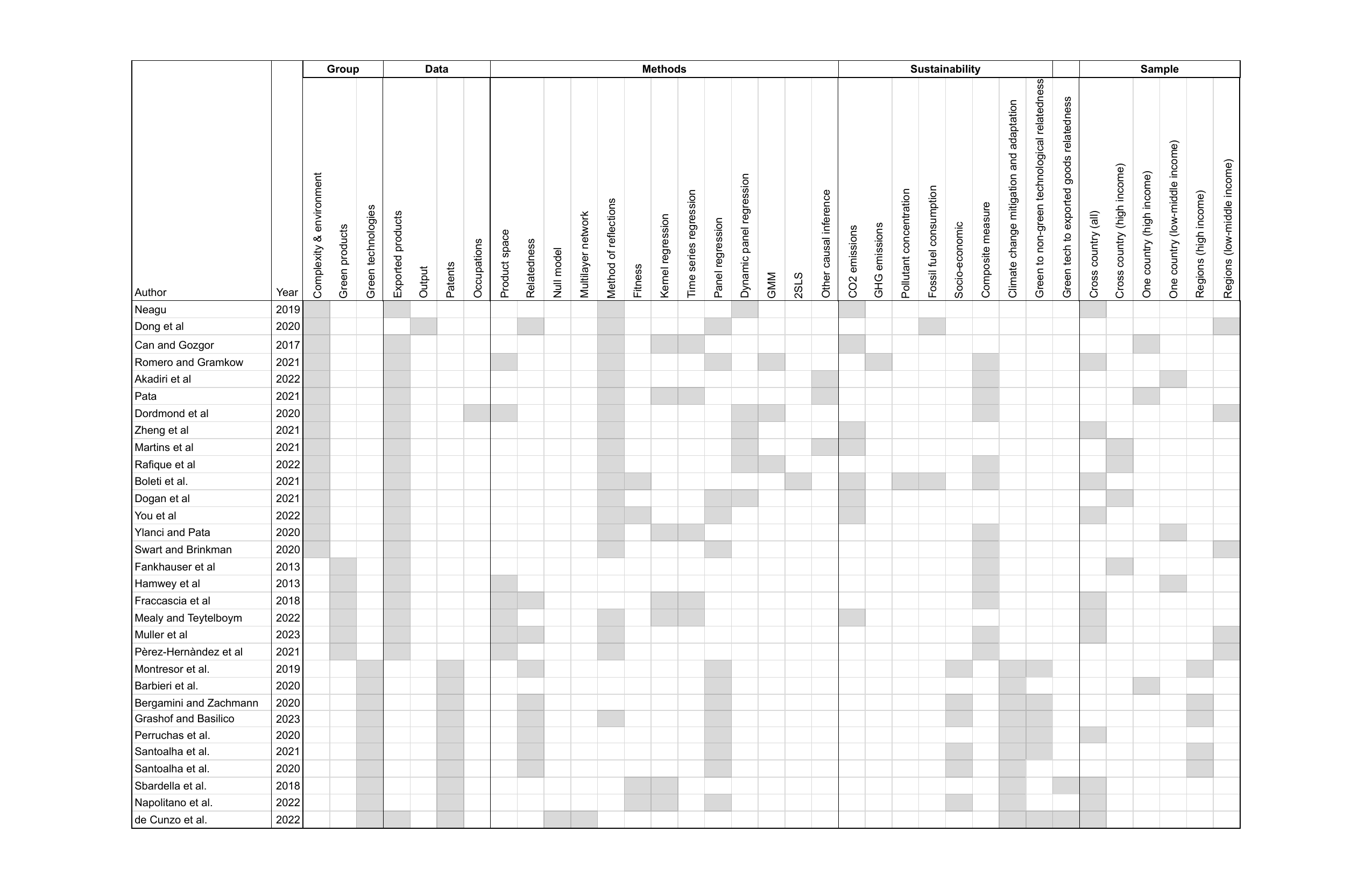}
    \caption{Summary of the reviewed articles, classified by comparable dimensions. 'Group' refers to the unit of analysis of the articles, the main feature considered in dividing empirical contributions in three different blocks, mirrored in the subsections that follow. 'Data' indicates the data sources employed; 'Methods' refers to the specific methodologies adopted to analyze the data, beyond the EC toolbox; 'Sustainability' refers to the main definition of environmental sustainability adopted by each paper; 'Sample' indicates the specific dataset used.}\label{fig:paper_matrix}
\end{figure}    
\end{landscape}

The rationale for the subdivision of the literature in three blocks follows rather naturally from the shared characteristics of the empirical applications of EC tools to study environmental sustainability. We divide the corpus of empirical literature based on the unit of analysis of the empirical study, namely i) geography (national or sub-national), ii) products and iii) technologies. A graphical representation of this representative corpus is summarized by Fig. \ref{fig:paper_matrix}, which identifies a number of features shared across the empirical contributions, namely: unit of observation (group), data, methods, definition of sustainability, and sample adopted. The following subsections reflect the organization of the first group of columns (group) in Fig. \ref{fig:paper_matrix}. 

The first subsection includes contributions that have explored the association between complexity metrics at the geographical level (mainly countries and regions) and variables related to environmental sustainability, such as aggregate CO$_2$ and GHG emissions. As can be inferred from Fig. \ref{fig:paper_matrix}, large share of this literature uses the Economic Complexity Index (computed using Eq. \ref{eq:reflections}) as explanatory variable, with a few exceptions that use instead the Economic Fitness and Complexity (Eq. \ref{eq:f-q}). The association between EC and environmental sustainability is analyzed in regression settings, which sometimes resort to panel dynamic techniques. Occasionally, causal claims are made by the authors. The sample of countries considered is rather heterogeneous, which hampers the rationalization of the findings. 

The second subsection focuses on the identification of green products, aiming to measure the green potential for green diversification based on export patterns. As shown in \ref{fig:paper_matrix}, the main tool used in this branch of studies is the product space. Their goal is to analyse the current status of countries' productive system from the point of view of green product. Most of these works do not try to asses relationship with target variables such as CO2 emission or other measures of sustainability. Mainly following \cite{mealy2022economic}, they use distances and proximity in the products space to construct a composite measure of 'greeness' of productive systems. Relying on export data, the geographical coverage is usually extended to most of the world, with few focuses on specific countries or regions.

The third subsection is constituted by empirical articles that use patent data to assess the readiness of regions for the green transition based on their existing capabilities examining the geographical patterns of green innovative capacity.
As can be appreciated from Fig. \ref{fig:paper_matrix}, this literature often explores the link and potential complementarity between green and non-green know-how, or productive capacity, mainly through proximity-based measures of relatedness between green and non-green technologies (Eq. \ref{eq_proximity}), with a few exceptions relying on an asset matrix-based definition of relatedness (Eq. \ref{eq:assit_matrix}), and exploring geographical patterns through the definition of Green and non-green technology based measures of Fitness (Eq. \ref{eq:f-q}).  These dimensions of relatedness or green technological complexity are used by different contributions as explanatory variables for a series of socioeconomic dimensions, such as inequality, digital literacy, or support for environmental policies.

\subsection{Economic Complexity and the environment}\label{eci-environ}

EC indices have been widely used by academics and policymakers to predict economic growth \citep{Hidalgo2009, Tacchella2012, Cristelli2013, Tacchella2018}. However, the pursuit of economic growth has been put under scrutiny by academics and society alike, due to its association to environmental degradation and climate change \citep{Raworth2017, IPCC2022}. Since the measurement of EC relies heavily on the nature of the products exported by countries or regions, which in turn impact the environment with different intensity (as they embody different levels of GHG emissions, have different energy requirements, and produce more or less polluting byproducts), there has been growing interest in understanding the relationship between the complexity of countries' productive structures and environmental impacts. 

An ever-growing body of literature (Table \ref{tab:ec_environment}) has investigated this relationship, looking especially at the export dimension of complexity and at country-level measures of environmental impacts. Contrarily to what early contributions seemed to suggest \citep{Hidalgo2021}, the last column in Table \ref{tab:ec_environment} hints towards aggregate findings that are far from unanimous, due to differing samples, geographical focus, and estimation techniques. This despite the fact that most of this evidence uses the Economic Complexity Index (Fig. \ref{fig:paper_matrix}) as explanatory variable; nevertheless, estimation techniques adopted to retrieve the effect of EC on the environment -- and the way in which environmental impacts are measured -- differ widely. 

One strand of this literature assumes a linear relationship between EC and environmental impacts, producing mixed evidence. In a rather comprehensive study, \citet{Romero2021} analyze the relationship between 67 countries' complexity levels and their CO$_2$ emissions, measured as aggregate, per capita, and product-specific. The latter is proxied by the Product Emission Intensity Index (PEI), which averages the emissions of countries exporting a product with comparative advantage -- following a methodology similar to \citet{Hartmann2017}. The study finds that lagged ECI is associated with a reduction of both emissions intensity and per capita, and that lower emissions are associated with more interconnected (complex) products. Looking exclusively at OECD countries, \citet{Dogan2021} also find a positive relationship between EC and the reduction of CO$_2$ emissions between 1990-2014. In addition, the authors show that the complexity of exports interacts positively with the consumption of renewable energy, contributing to mitigating environmental impacts in high income countries. One possible underlying mechanism is the emergence of greener occupations -- and therefore greener economic activity -- in regions with higher complexity, although the spatial dynamics of evolution towards the green occupations space prove to be rather sticky \citep{Dordmond2021}.

Although the studies mentioned above seem to agree on the positive contribution of EC to the environment, comparing their results is not always straightforward. For instance, \citet{Boleti2021} rely on a measure of environmental performance -- the Environmental Performance Index\footnote{ The Environmental Performance Index used in this study is obtained aggregating 40 indicators across 11 issues, such as Climate Change Mitigation; Air Quality; Sanitation and Drinking Water; Heavy Metals; Waste Management; Biodiversity and Habitat; Ecosystem Services; Fisheries; Acid Rain; Agriculture; Water Resources. For more information, see the latest EPI report: \url{https://epi.yale.edu/downloads/epi2022report06062022.pdf}} -- to show that increased complexity is associated with better environmental performance across 88 low- and high-income countries over a (short) period spanning between 2002 and 2012. However, in the same period, higher complexity is also associated with worse air quality\footnote{ Air quality is a composite index used to compute the EPI, and which in turn aggregates different indicators that quantify the number of age-standardized disability adjusted life-years lost per 100,000 persons (DALY rate) due to exposure to: PM$_{2.5}$ exposure, household solid fuels, ozone exposure. Moreover, the air quality indicator includes measures of population-weighted annual average concentration at the ground level for: NO$_x$, sulfur dioxide, carbon monoxide, and volatile organic compound.}, higher emissions of CO$_2$ (kg per 2011 PPP USD of GDP), methane (kg of CO$_2$ equivalent), and nitrous oxide (thousand metric tons of CO$_2$ equivalent).  

Some studies also find that EC is linked to an increase in CO$_2$ emissions. For instance, \citet{Neagu2019a} examine the heterogeneous effects of EC on GHG emissions in European countries. Their work shows that higher EC is positively linked to the growth of GHG emissions by countries, although this happens faster for countries with relatively lower levels of EC. These results are consistent with other studies focusing exclusively on the most complex countries: using time-series estimation techniques, \citet{Martins2021} find that higher EC is positively associated, in an unidirectional way, the levels of CO$_2$ emissions in the top 7 countries in the EC ranking. Nevertheless, integration in international trade has mitigated such negative effects, allowing early industrializers to shift towards knowledge-intensive, less polluting tasks. Similarly, \citet{Rafique2022} resort to dynamic panel data estimation techniques to find that ECI -- along with urbanization and export growth -- is positively linked to the ecological footprint\footnote{The authors rely on an index -- the Ecological Footprint Index -- which includes factors such as area occupied by forests, cropland, grazing, built-up land, fishing, and CO$_2$ emissions.} in the top 10 complex countries. 

The mixed evidence on the relationship between EC and environmental sustainability could also be explained by the fact that such linkage may be non-linear. It has been argued that countries increase their environmental impact -- in terms of GHG and pollutant emissions -- as they industrialize, reaching a peak in their emissions per capita. However, as they move towards more sophisticated activities and services, emissions per capita start to decrease, while GDP keeps growing. The reversed U-shape of the relationship between GDP and emissions has been named Environmental Kuznets Curve (EKC) \citep{Grossman1991, Selden1994, Grossman1995}, building on the work by \citet{Kuznets1955}, who observed the same reversed U-shaped relationship between inequality and economic growth unfolding along with the process of structural transformation\footnote{ The EKC hypothesis suggests that the link between economic growth and its environmental impact depends on two channels: the change in composition of economic activities, and in the techniques of production \citep{Grossman1991}. As countries shift from agricultural to industrial activities, there is a rise in carbon-intensive economic activities, leading to a positive correlation between GDP per capita and emissions. Thereafter, as modern techniques of production are introduced -- leading to more efficient energy use and a shift to less energy-intensive services -- the relationship between GDP per capita and emissions becomes negative. This results in a reversed U-shaped relationship between GDP per capita and emissions.}. 
A growing body of literature has attempted to identify a complexity-based EKC; the intuition behind a reversed U-shaped relationship between EC and CO$_2$ emissions lies in the fact that, as countries become 'fitter' and accumulate productive capabilities, they shift their specialization towards more knowledge-intensive goods, with the latter being greener. 

An ECI-based EKC has been identified especially for high-income countries, which find themselves at a mature post-industrial stage -- that is, beyond the peak of emissions that coincided with higher intensity of employment manufacturing industries. In particular, ECI-based EKCs have been empirically tested for France \citep{Can2017} and the US \citep{Pata2021}, as well as for a sample of 25 European countries \citep{Neagu2019} and for leading exporters \citep{Zheng2021}. However, the same cannot be said for emerging economies, as they are still intensive in manufacturing or extractive industries, which are associated with higher GHG emissions. For instance, empirical tests of the EKC in China -- both at the aggregate \citep{Yilanci2020} and regional level \citep{Akadiri2022} -- find no evidence of a reversed U-shaped relationship between complexity and CO$_2$ emissions. On the contrary, higher complexity appears to be associated with higher CO$_2$ emissions in both cases, despite the negative association between regional complexity and coal consumption in China \citep{Dong2020}. In the case of Brazilian regions \citep{Swart2020}, the EKC hypothesis is validated using a quadratic fit between ECI and several indicators of environmental quality. The hypothesis holds for waste generation, but not for forest fires, deforestation, and air pollution. It is however important to distinguish that, in this case, the EKC tested does not refer to carbon emissions -- which have global consequences on climate change -- but to factors related to local environmental degradation. 

It is important to mention that both the theoretical and empirical grounds of the EKC hypothesis -- as well as its normative implications \citep{Savona2019} -- are not uncontroversial. First, the future ability of emerging countries -- like China and Brazil -- to decouple economic growth from GHG emissions is all but certain. This casts a doubt on whether we can expect low- and middle-income countries to go down the same path as currently high-income countries. Even if this will be the case, it must be taken into account that the polluting industries in which countries were once specialized may be simply outsourced to other countries, lending support to the pollution haven hypothesis\footnote{The pollution haven hypothesis states that regulations in high-income countries aiming at reducing GHG emissions may lead firm to relocate in countries with looser environmental regulations, thus leading to an outsourcing of CO$_2$-intensive activities rather to their effective reduction worldwide. For a case linking EC to the pollution haven hypothesis, see \citet{Dong2020}.} \citep{Cole2004}. Moreover, the right part of the EKC -- where the decoupling of economic growth and CO$_2$ emissions is observed -- is populated only by a few developed countries \citep{Csereklyei2016}. Even if the same relationship was to hold also for developing countries in the future, waiting for all countries to move beyond the plateau may irremediably compromise the environment, as highlighted by the IPCC \citeyear{IPCC2022}.  

Moreover, as noted by \citet{Stern2004} in his critical review, the EKC literature tends to tread GDP per capita as an exogenous variable, ruling out any possible feedback from the environmental dimension to the productive sphere. Additionally, focusing on individual pollutants or GHG emissions has neglected the changing composition of emissions associated to the change in the productive structure. More recently, it has been noted that convergence approaches may be more relevant in describing the relationship between economic development and environmental impact, accounting also for non-industrial GHG emissions \citep{Stern2017}. 

Another issue worth discussing revolves around the data used in the empirical analyses discussed above. While most of the empirical literature reviewed in this section relies on exported products data, the trade dimension is not the only relevant aspect in terms of environmental sustainability. As shown by \citet{Stojkoski2023}, the CO$_2$ intensity across countries is explained more fully by a combination of trade, technological, and scientific complexity, computed respectively using data on exported products, patent applications, and scientific publications. Moreover, the authors show that not all complexity algorithms work equally well in predicting CO$_2$ intensity, showing that the Fitness algorithm \citep{Tacchella2012} outperforms competing measures. As shown in Fig. \ref{fig:paper_matrix}, however, only a small minority of authors opts for the EFC algorithm (in this review, \citealt{Boleti2021, You2022}). 

Finally, we would like to raise some methodological issues regarding the evidence canvassed in this section. Apart from rare exceptions (i.e \citealt{Can2017}), the empirical papers that examine the relationship between EC and indices of environmental performance, use a measure -- the Economic Complexity Index \citep{Hidalgo2009} -- which should only be treated as relative within each year (as explained in Section \ref{sec:compl_measures}). As the term of reference for the measure changes every year, changes in ECI over time do not have any longitudinal interpretation, as the scale with which the index is measured changes every year. In order to address this, the use of EC metrics within longitudinal regressions should rely on the projection of product complexity in a given year of the series, upon countries' Revealed Comparative advantages in every year \citep{sbardella2018green, Operti2018} -- i.e. an ''exogenous'' complexity, as explained at the end of Section \ref{sec:compl_measures}. Alternatively, comparability over time is ensured by measures that maintains an invariance of scale, as suggested by \citet{mazzilli2024equivalence} and explained in this article after Eq. \ref{eq:f-q}. 

One of the advantages offered by complexity methods in studying the sustainability transition is that these allow to observe productive structures and trajectories at a finely disaggregated level, for instance identifying single products or technologies towards which diversification should be steered in order to facilitate the transition towards greener activities. The role of products and technologies will be discussed in the following two subsections. 




\input{tab_ec_environment_sht.tex}


\subsection{Economic Complexity for assessing green productive capabilities}\label{green-prod}

Product-level data is the most commonly used data source in the EC framework. However, it is less frequently employed in studies linking EC and the sustainability transition, as compared to data on green patenting activity. This is mainly due to the difficulty of defining the 'green products' outlined in Section \ref{data}. 
Table~\ref{tab:green_prod} summarizes our selection of articles for this line of research.

Few attempts have been made to assess the green complexity/potential of national industrial systems. So far, the literature focused on the following research questions: Are green products more complex? How related to new green products is a given productive structure? Is there a significant difference in the dynamics of the product space for green and non-green products?
Thus, on the one hand, scholars have tried to characterize green products as a particular subset of products in international trade networks by using standard EC tools. On the other hand, the main question -- 'how prepared is a country to a green transition?' -- is tackled mainly via network tools such as the products space, not introducing specific tools or frameworks for green commodities. Most contributions in the literature in fact appliy the standard EC methodology and restrict the analysis to the subset of green products. 

For instance, \cite{fankhauser2013will} assess the starting point of the green race for eight countries (China, Germany, UK, USA, France, Italy, Japan, and South Korea) for the period 2005-2007. They analyze three main factors of the green transition: green conversion, estimated using the Green Innovation Index (GII), an index constructed using patent data; change in RCAs; and green production at the outset, which, due to lack of specific data on this dimension, is assumed to be proportional to total production. Their findings reveal different areas of competitiveness among countries, detected from a correlation analysis of their Green Innovation Index and changes in RCAs. They find no actual correlation between patenting GII and RCA competitiveness, which could mean a lack of policies for the green conversion in most of these countries.

\cite{hamwey2013mapping} apply the product space technique to identify opportunities for countries in green production by taking into consideration an arbitrary set of eleven environmental groups of products (in the SITC 4 classification at 4 digits) drawn from  the list proposed by the WTO Committee on Trade and Environment (CTE) in 2011. In particular, their analysis focuses on Brazil, finding few selected products as potential opportunity of green diversification. However, the authors recognize the weakness of the green product definition, classification, and selection, as well as the limitation of neglecting the dynamics of the RCA time series they use to construct the products space.

Building on \cite{hamwey2013mapping}, \cite{fraccascia2018green} apply the product space description for green products and measure the proximity between green products -- as defined by the environmental goods and service sector classification by EUROSTAT (2009) -- and non-green products in which world countries display an RCA>1. Using panel regression analysis, they find this proximity to be significant for the development of the export of green products in a 4-years horizon.

In \cite{mealy2022economic} two lists of green products are defined: a broader list comprising 293 green products, obtained by merging the WTO Core list, OECD lists, and the APEC list, and a shorter list of 57 renewable energy products. Using all products available in COMTRADE, they first compute the PCI of all products and construct a product space. Then, they focus on green products extracting the Green Complexity Index (GCI), the sum of green products' complexity, and the Green Complexity Potential (GCP), estimated by the proximity to green products in the product space. They reveal the complexity of green products to be significantly higher than the entire products baskets. Notably, they find a negative correlation between GCI and CO$^2$ emissions per capita, as well as a positive correlation between GCP and an increase of export in green products. 

\cite{perez2021mapping} follow \citet{mealy2022economic} and apply GCP and GCI at a sub-national level, studying the green potential of Mexican regions. They use an adapted version of the CLEG classification to characterize the green product space of regions. While their findings confirm an overall positive correlation between GCI and GCP, some interesting outliers arise are highlighted. For example, the state of Jalisco shows a high GCP and a comparative advantage on few high complexity products, but a low diversification in green products. This may be explained by some geographical or fine-grained factors, not captured in these indexes, that are less important at larger geographical scale.

\cite{muller2023economic} apply the product space methodology focusing on the production of green hydrogen. They define a list of 36 goods 'that are potentially relevant for the production of green hydrogen from solar and wind power in a stand-alone configuration'. They also define a larger list by selecting 434 products with a high proximity, in the product space, to the first 36 goods. In agreement with the literature, they find higher PCI of products in these two lists compared to the entire HS products classification. Then they focus their study on European and Middle East and North African (MENA) countries, analyzing how their GCI evolved in the period 1995-2019 and assessing their readiness for the upcoming increasing demand of green hydrogen, and their potential role in this specific supply chain.

\input{tab_green_prod.tex}

\subsection{Economic Complexity for assessing green technological capabilities}\label{green-tech}

In recent years, the EC literature exploring the intricate relationship between green technological development, regional specialization dynamics, and policies has grown substantially. By drawing insights from evolutionary economic geography, these studies have revealed valuable insights into the dynamics and drivers of green technology advancement. This has been pursued by looking at the preparedness of European NUTS2 regions or national economic structures to the green technology race, and by highlighting the potential complementarities between green and non-green knowledge bases and/or productive structures. 

The majority of the papers reviewed in this section rely on climate change mitigation and adaptation patent data as a proxy for low-carbon innovation, using mainly the \textsc{Env-Tech} classification and the Y02/Y04S tagging scheme presented in Section \ref{data}. This is increasingly becoming the gold standard to measure green innovative activities. While the limits of using patent data for the study of technology development should be acknowledged (see \textit{e.g.}, \citealp{arts2013inventions, griliches1998patent, lanjouw1998count}), information on patents is widely available, and it can provide an array of quantitative information on the nature of inventions and their applicants or inventors, including their geographical location, allowing to easily geo-localise patents both at country and local levels \citep{dechezlepretre2011invention}. Moreover, patent data can be disaggregated into increasingly fine-grained technological areas, allowing very specific green technologies to be identified \citep{hascic2015measuring}.  
This granularity is particularly helpful in the use of EC techniques to study technological specialization \citep{boschma2015relatedness,napolitano2018technology,pugliese2019unfolding}.
The application of EC approaches to reveal technological advantages in each technology field rests on the idea that the criteria to assign patent applications to specific domains are based on the identifying characteristics of the expertise that is necessary to introduce successful inventions. As a matter of fact, complex technologies appear almost exclusively in the portfolios of high-complexity countries and less diversified countries operating in less complex sectors.

As can be appreciated in Table \ref{tab:green_tech}, where we summarize the main results, methodologies, and data employed in the papers included in this section, some of these contributions focus on the interplay between green capabilities and the maturity of green technologies in shaping the way in which the technological portfolios of regions or countries grow and evolve over time. Other contributions explore instead the relation between income inequality and innovative capacities (both green and non-green) at the country level, revealing the influence of socioeconomic and policy factors (e.g. smart specialization strategies, political support, digital literacy) . 
Overall, these contributions offer a cohesive narrative, that emphasizes the important role played by local capabilities and relatedness to non-green knowledgge bases or productive capacity in fostering new green technological innovation, providing valuable guidance for policymakers seeking to promote sustainable models of economic development.

\cite{perruchas2020specialisation} investigate the relationship between country characteristics and knowledge structures in the progression of green \textsc{Env-Tech} patenting, and national specialization or diversification patterns from the end of the 1970s. The authors focus on the life cycle of green technologies and propose a 'ladder of green technology development'. Their evidence emphasizes that not only countries diversify towards green technologies related to their existing competencies, but also that green specialization follows a cumulative path towards more mature technologies. Technology maturity appears to be more relevant than a country’s economic development, while technology complexity -- computed through the ECI algorithm introduced in Eq. \ref{eq:reflections} applied to green patent data -- does not prevent further specialization. By focusing on a panel of US states from the early 1980s, the same authors \citep{barbieri2020specialization} examine the role of related/unrelated variety in green technology development and its heterogeneous effect over the technology life cycle. 
The authors observe that unrelated variety plays a positive role in fostering green innovation at early stages of the technology life cycle, while as technologies mature, related variety becomes a more important driver. 

\cite{sbardella2018green} introduce Green Technological Fitness (GTF), a measure of green innovative potential based on green patenting as defined in Eq. \ref{eq:f-q} -- here identified within PATSTAT through the \textsc{Env-Tech} catalogue. Taking a geographical approach, the authors identify heterogeneous global patterns in green technological competitiveness, with the United States, France, and Germany as stable leaders; Eastern and Southern European countries gradually gaining importance; and East Asian countries starting from the periphery and rapidly establishing themselves as key actors. Finally, by analyzing the distribution of countries’ innovation capacity across areas of specialization, they document that innovation in green technology has become more horizontal, with bigger efforts being observed in cross-domain, or enabling technologies.

Finally, focusing only on the technological dimension and not looking at geographical dynamics, \cite{de2022trickle} explore the connection between green technological innovation capacity and productive capabilities. In order to do so, the authors define a statistically validated network -- as defined in Eq. \ref{eq:assit_matrix} -- connecting comparative advantages in Y02 green technologies to the contemporaneous or subsequent comparative advantages in exported products. When looking at same-year co-occurrences of single products at technologies, their findings emphasize a large number of significant links between green technologies and raw materials, especially critical minerals such as cobalt. In contrast, when selecting ten-year time-lagged green technology-product links, they suggest that green technology is better integrated into manufacturing, and that more complex spillover effects emerge with new additional links of products and technologies with higher complexity. 

Evidence of different dimensions of complementarity between green technologies and non-green pre-existing innovative capacity in European NUTS2 regions is investigated by five articles \citep{bergamini2020exploring,montresor2020green, barbieri2022regional,sbardella2022regional,grashof2023dark} using different approaches to assess relatedness. 
Firstly, \cite{bergamini2020exploring} use regional patent data for Europe sourced from the REGPAT database \citep{maraut2008oecd} to predict the potential of European regions to acquire (or maintain) competitive advantages in developing green technologies. To this aim, the authors first leverage a network-based measure of relatedness (as in Eq. \ref{eq_proximity}) between non-green technologies to estimate the potential technological advantage (RTA) in green technologies of each region. In a second stage, they identify a set of socioeconomic variables that hold a statistically significant association with observed/estimated RTAs and derive possible policy implications.

Secondly, \cite{montresor2020green} explore the relationship between \textsc{Env-Tech} green patenting and smart specialization strategies in European regions, paying attention to the role of key enabling technologies.\footnote{ The term key enabling technologies denote six crucial technological domains identified by the European Commission: industrial biotechnology, nanotechnology, micro- and nanoelectronics, photonics, advanced materials, and advanced manufacturing technologies.} With the aim of understanding if the development or acquisition of new sustainable technologies is related to regional technological innovative capacity, the study shows that there is a strong connection between green technologies and both green and non-green pre-existing knowledge bases. Moreover, key enabling technologies support the shift towards green technologies also mitigating the impact of relatedness to pre-existing technologies. 

Thirdly, \cite{barbieri2022regional} and \cite{sbardella2022regional} investigate the nexus between non-green (A-H PATSTAT patents) and green (Y02-Y04S) innovative capacity in and the green development potential of European regions, computing measures of \textit{exogenous} fitness as introduced in Section \ref{sec:compl_measures}, regional Non-Green Technology Fitness (NGTF), and Green Technology Fitness (GTF). 
Using the statistically validated network approach introduced in  Section \ref{sec:relatedness}, they define a green potential metric quantifying the relatedness (as defined in Eq. \ref{eq:assit_matrix}) between non-green and green regional knowledge bases. The authors document a heterogeneous but stable distribution of non-green and green innovative competitiveness, and reveal a degree of complementarity between non-green and green knowledge capabilities, albeit dependent on the patent portfolio composition -- especially for regions that have not fully developed the entire set of Y02-Y04S technologies. 

Finally, by investigating the nexus between EU NUTS2 regions’s knowledge base, economic performance, and green technology specialization patterns (identified through WIPO IPC Green Inventory) between 2000 and 2017, \cite{grashof2023dark} argue that the sustainability transition will not widen regional disparities. Their analysis, where a proximity-based relatedness is computed according to Eq. \ref{eq_proximity}, suggests that both high- and low-income regions can diversify into green technologies. While high-income regions are more specialized in technologies related to green innovations, they do not necessarily have more entries in green technologies. Instead, related technological capabilities appear to be more important for green diversification than economic strength. Notably, low-income regions with high green relatedness can successfully diversify into new, and even more complex, green technologies. 

Different contributions in this field are in agreement in suggesting that relatedness is a driving force behind diversification in green technologies. However, this new field of analysis has not yet fully emphasized the role of the socioeconomic fabric and institutional set-up in sustaining the sustainability transition. 
The sustainability transition literature has paid greater attention to the policy and socioeconomic dimensions, however, putting forward an important but not systematic collection of case studies, has often failed to provide generalizable or scalable evidence on the role of environmental policies or local characteristics in shifting towards a more sustainable economy. The three following articles go in this direction and analyze how green technological competitiveness, diversification in green technologies, and/or relatedness to non-green knowledge interact with a number of socioeconomic characteristics of countries or regions.

Looking at the regional dimension, \cite{santoalha2021diversifying} examines the relationship between political support for environmental policies and the diversification of green technologies in European regions. By employing a  measure of relatedness similar to the one encapsulated in Eq. \ref{eq_proximity}, they find evidence of a stronger relationship between related capabilities and green diversification, as compared to regional policies. 
However, national political support appears to mitigate the importance of capabilities.

Building on a similar metric for relatedness, \cite{santoalha2021digital} operationalize the notion of capabilities by studying whether and to what extent digital literacy -- as a proxy for the competencies embedded in ICT infrastructures -- foster diversification in green and non-green technologies across European regions. The level of digital skills in the workforce has a positive impact on a region's ability to specialize in new technologies, especially in green domains, with e-skills moderating the effect of relatedness.

Shifting to a country-level analysis, \cite{napolitano2022green} investigate whether and to what extent income inequality is a barrier to a country’s environmental innovative capacity, proxied by a measure of Green Technological Fitness based on \textsc{Env-Tech} technologies. To this end, and differently from \cite{sbardella2018green}, they define a measure of \textit{sectoral} fitness, to provide a more realistic assessment of green technology complexity: first, they account for the full technological spectrum in computing the EFC algorithm (Eq. \ref{eq:f-q}); then, they select only the complexities of \textsc{Env-Tech} technologies in computing GTF. A negative and significant relationship between income inequality and GTF is observed; by contrast, no significant association is found when all technologies are considered. However, while for high-income countries inequality does not appear to be a barrier, there is an income threshold below which it is unlikely to develop a sufficient number of complex technologies to achieve high green fitness. Low inequality reduces such thresholds allowing middle-income countries to achieve greater green innovative capacity.

\section{Conclusions}\label{conclusions}

In this review we have provided an encompassing account of the empirical literature implementing EC methods and metrics to understand the sustainability transition. First, the paper has summarized the most used product- and patent-level data sources to compute EC metrics, with a particular focus on green economic activities. Secondly, the review has attempted to harmonize the most relevant methods adopted in the literature on EC and environmental sustainability, in order to identify similarities and differences across methods relying on a common framework and language. Third, the three main blocks of empirical literature linking EC methods and metrics have been reviewed, looking at i) the relationship between countries' and regions' EC and the environment, and the role of ii) green products and iii) green technologies in fostering the sustainability transition.

The growing literature on EC and the environment suggests that EC approaches can be a useful lens to better understand how productive structures and technological capabilities can be steered into the sustainability transition. The empirical literature reviewed here offers some evidence that can be summarized as follows. With respect to the first block of literature, which examines how trade-based EC measures link to environmental outcomes such as GHG emissions, ecological footprint, and environmental degradation, we conclude that the evidence produced is mixed. On the one hand, high-income countries -- which are more specialized in knowledge-intensive activities -- seem to exhibit a positive association between the complexity of their export basket and their ability to preserve the environment. However, the same relationship cannot be observed in less mature, emerging economies, where increasing complexity is associated to higher environmental impact. Empirical test of the EKC hypothesis have shown that countries become able to decouple economic growth from carbon emissions only after passing the industrialization phase.
Having this said, the extant literature on green EC has some important limitations. First, the literature on country/regional complexity and sustainability relies largely on aggregate measures of EC. This approach can be limiting if confined to examining the relationship between complexity and environmental variables, without taking advantage of the high level of disaggregation made possible by EC methods. Secondly, as we have argued in Section \ref{eci-environ}, the quest for empirical validation of the EKC hypothesis can be of little normative utility, considering that not all countries will be able to emulate the diversification and specialization patterns of today's high-income countries, some of which are specialized in knowledge-intensive activities. Moreover, as shown by the \citet{IPCC2022}, the current efforts to curb carbon emissions also in high-income countries may be insufficient to prevent a climatic catastrophe. 

Looking at the second block of literature, one of the main goals of these studies is to steer production systems toward a green, feasible, and just transition. While the nature of the trade-off between these elements is left to more theoretical studies, EC has the potential to identify, characterize, and measure potential paths of such transformation. From the reviewed attempts, it is clear that a greener production of commodities requires higher capabilities. Green products show higher complexity and are intertwined to non-green production. As stressed in the Data section of this paper, the studies relying on green classifications for products are subject to the problem of how to define a green product: different classifications based on inconsistent definitions may undermine the replicability of the results. A green sub-classification in the Harmonized System would be desirable for the upcoming updates, so to have a single and exhaustive database for green products. On the methodological front, most contributions in the EC literature studying green products apply the well known product space description to approach the problem of identifying the green potential of productive structures. While the product space is indeed useful to visualize the distances between the current production and green products, it has been shown that this approach has low accuracy in quantitative forecasting of future diversification \citep{tacchella2023relatedness}.

The third block of studies on green technological development yields several key conclusions. Firstly, there is evidence of countries engaging in a dual strategy of diversification and specialization, entering green technologies aligned with existing competencies while specializing in mature technologies with accumulated experience. 

\begin{landscape}
\input{tab_green_tech.tex}

\end{landscape}

Geographically, there is a mix of stable leaders and emerging players in green technological competitiveness, with Eastern and Southern European countries gaining prominence. Furthermore, economic factors --  such as income inequality -- affect a country's environmental innovative capacity. For instance, lower income inequality in middle-income countries seems to lower barriers to the successful development of complex green technologies. Finally, the complementarity between non-green and green innovation capabilities, the influence of regional factors and policies, and the significance of digital literacy and e-skills in promoting green technology adoption are highlighted. These findings provide valuable insights into the complex dynamics of green technological development, emphasizing the interplay between capabilities, regional factors, policies, and socioeconomic aspects.

Whilst the studies on technology-based EC can shed light on various aspects of green technological development, it is important to acknowledge their limitations. Firstly, the analyses often rely on patent data as a proxy for technological innovation, which may not capture the full spectrum of green technologies, or account for innovations that are not patented. This could lead to a potential under-representation of certain sectors or technologies. Secondly, the studies primarily focus on regional or national levels of analysis, which may overlook the dynamics at the firm or individual levels. The role of specific organizations or entrepreneurs in driving green innovation is not extensively explored, potentially limiting the understanding of micro-level factors influencing technological development. Lastly, the studies primarily analyze the relationships between different variables and identify correlations rather than establishing causal relationships. While the associations observed are informative, further research is needed to delve deeper into the underlying mechanisms and causality. These limitations highlight the need for further research that encompasses a broader range of data sources, considers multiple levels of analysis, and employs rigorous methodologies to better understand the complexities of green technological development and its implications for environmental sustainability.


The discussion on the empirical literature presented in Section \ref{review} uncovers a number of areas that require further exploration. First, the productive and technological structures of countries are always accounted for separately in relation to their linkage with environmental technologies. Whilst there have been previous efforts linking the productive, technological and scientific capabilities of countries with each other \cite{pugliese2019unfolding}, and one contribution focused on the green know-how-productive capacity nexus \cite{de2022trickle}, such effort should be extended to better understand which specific capabilities are the most conducive to green specialization across fields. 
Second, and related to the previous point, scientific capabilities have not yet been included in the discourse on the sustainability transition from a complexity perspective. Their relationship with the development of green capabilities should be examined in depth.  In fact, being cognizant of the broad consensus on the key role of  green technologies for the sustainability transition, these cannot be viewed as a one-size fits all solution, and should be integrated in a broad policy agenda targeting wider dimensions of socioeconomic sustainability \citep{parkinson2010coming,sarewitz2008three}.

Third, ensuring a green and fair transition will require to take into account its interrelation with the process of structural change, labour reallocation, and their distributive consequences. Future research will have to assess the job creation/destruction potential of the green transition, and to identify the transversal skills required to ensure a seamless reallocation of workers across jobs. By the same token, not all territories and geographies are equally well equipped with the necessary knowledge base to diversify away from carbon-intensive production and technologies. It will be paramount to identify the most viable ways for regions and countries to enter green economic activities, bearing in mind the potentially negative (or positive) impact on local labor markets.

Finally, green technologies and products may create further pressure on the environment due to their dependency on natural commodities and minerals \citep{Marin2021}, such as rare earth elements, lithium, cobalt, and others \citep{de2023mapping}. Precisely, the World Bank \citep{Worldbank2020} estimates that meeting the 2°C scenario by 2050 for energy storage alone will require a 450 percent increase in the production of graphite, lithium, and cobalt. Therefore, while pursuing the green transition may contribute to reducing global dependence on fossil fuels, keeping up with current levels of energy demand will shift the pressure towards the production and trade of raw materials, neither of which is exempt from complications. 
For this reason, future research will need to examine the production processes and recycling potential associated to each green product, the CO$_2$ production incorporated in each value chain, its raw material content, the safety (also concerning toxicity and pollutant exposure) and workplace condition of the labor force employed in its production, and the environmental impact of production and use (e.g. life cycle emissions, energy content, and waste management).

\section*{Acknowledgments}

DM, AP, and AS acknowledge the financial support under the National Recovery and Resilience Plan (NRRP), Mission 4, Component 2, Investment 1.1, Call for tender No. 104 published on 2.2.2022 by the Italian Ministry of University and Research (MUR), funded by the European Union – NextGenerationEU– Project Title  "WECARE" – CUP 20223W2JKJ by the Italian Ministry of Ministry of University and Research (MUR).
DM, AP, and AS acknowledge the financial support under the National Recovery and Resilience Plan (NRRP), Mission 4, Component 2, Investment 1.1, Call for tender No. 1409 published on 14.9.2022 by the Italian Ministry of University and Research (MUR), funded by the European Union – NextGenerationEU– Project Title "Triple T – Tackling a just Twin Transition: a complexity approach to the geography of capabilities, labour markets and inequalities" – CUP F53D23010800001 by the Italian Ministry of Ministry of University and Research (MUR).

\bibliographystyle{apalike}
\bibliography{review-green}

\newpage

\appendix
\section*{Appendix}\label{sec:appendix}
\input{appendix-patent_data}

\end{document}

%% file: section-patent_data.tex
The main data sources for green or non-green patenting activity at the national or local level are the European Patent Office's (EPO) PATSTAT and the OECD's REGPAT.  PATSTAT is a comprehensive database covering patent applications filed in more than 70 national and international patent offices, including the most important ones -- e.g. the EU, US, and Japan patent offices. 
REGPAT is instead published annually by the OECD and covers the subset of PATSTAT patent applications filed at the EPO since its inception in 1978.  While PATSTAT covers a wider set of patent documents and a longer time-span, REGPAT covers EPO's applications, that are high-quality and extremely reliable -- we refer to the paper's Appendix \ref{sec:appendix} for a more detailed comparison between these two data sources.

The technological fields associated to the patents recorded in PATSTAT and REGPAT follow two classifications: the International Patent Classification (IPC) and the Cooperative Patent Classification (CPC). 
Both classifications follow a similar hierarchical structure, spanning from codes with a very detailed description (e.g., G02B 1/02: optical elements [\dots] made of crystals, e.g. rock-salt) to codes aggregating many detailed technologies under a broader common technological area (e.g., G02: optics; G: physics).
For example, at 4-digits, the classifications have around 600 unique codes, while at 8-digits there are approximately 7000 codes in the IPC and 10000 in the CPC.  
Despite the strong similarities between the two classifications, only the CPC has a section dedicated to Climate Change Mitigation and Adaptation Technologies (CCMTs), the 'Y02/Y04S' tagging scheme, nested under section Y.
Table \ref{tab:tech_sections} reports the 1-digit codes and the titles of the sections comprising the IPC and CPC classifications. 

\begin{table}[bh!]
    \centering
    \caption{Sections of the IPC and CPC classifications.\textsuperscript{\textdagger}\\}
    \label{tab:tech_sections}
    \resizebox{.9\textwidth}{!}{%
    \begin{tabular}{c l c c}
        \toprule
         \textbf{Code} & \textbf{Title} & \textbf{IPC} & \textbf{CPC}\\
         \toprule
         A & Human necessities & Y & Y \\
         B & Performing operations; Transforming & Y & Y \\
         C & Chemistry; Metallurgy & Y & Y \\
         D & Textiles; Paper & Y & Y \\
         E & Fixed constructions & Y & Y \\
         F & Mechanical engineering and weapons & Y & Y \\
         G & Physics & Y & Y \\
         H & Electricity & Y & Y \\
     \textbf{Y} & \textbf{New technologies, including CCMT\textsuperscript{\textdaggerdbl} \& Smart Grids} & \textbf{N} & \textbf{Y} \\
        \toprule
    \end{tabular}
    }
    \\
    \justify{
    \footnotesize\textsuperscript{\textdaggerdbl}Climate Change Mitigation and Adaptation Technology.}
    
\end{table}




\begin{table}[hptb]
\centering
\caption{CPC Y02/Y04S (2018) tagging scheme \citep{epoY}.}
\label{tab:1}
\resizebox{.98\textwidth}{!}{%
\begin{tabular}{@{}cl@{}}
\toprule
\multicolumn{1}{l}{\textbf{Class or Subclass}} & \multicolumn{1}{l}{\textbf{Title and description}} \\ \midrule
\textbf{Y02} & \begin{tabular}[c]{@{}l@{}} \textbf{TECHNOLOGIES  OR APPLICATIONS FOR MITIGATION OR ADAPTATION} \\ \textbf{AGAINST CLIMATE CHANGE}\end{tabular} \\  \midrule
Y02A & Technologies   for adaptation to climate change \\ \\
Y02B & \begin{tabular}[c]{@{}l@{}}Climate change mitigation technologies related to buildings, e.g. housing, house appliances or \\ related end-user applications, including the residential sector\end{tabular} \\ \\
Y02C & Capture, storage,   sequestration or disposal of greenhouse gases \\ \\
Y02D & \begin{tabular}[c]{@{}l@{}}Climate change   mitigation technologies in information and communication technologies,\\  i.e. information and communication technologies aiming at the reduction of their own energy use\end{tabular} \\ \\
Y02E & \begin{tabular}[c]{@{}l@{}}Reduction of   Greenhouse Gases (GHG) emissions, related to energy generation, \\ transmission or   distribution, including renewable energy, efficient combustion, \\ biofuels, efficient transmission and distribution, energy storage, and hydrogen technology\end{tabular} \\ \\
Y02P & Climate change   mitigation technologies in the production or processing of goods \\ \\
Y02T & Climate change   mitigation technologies related to transportation, e.g. hybrid vehicles \\ \\
Y02W & Climate change   mitigation technologies related to wastewater treatment or waste management \\ \\\midrule
\textbf{Y04} & \begin{tabular}[c]{@{}l@{}} \textbf{INFORMATION   OR COMMUNICATION TECHNOLOGIES HAVING AN IMPACT }  \\ \textbf{ON OTHER TECHNOLOGY AREAS}\end{tabular}\\\midrule
Y04S & \begin{tabular}[c]{@{}l@{}}Systems integrating technologies related to power network operation,  communication or\\ information technologies for improving the electrical power generation, transmission, \\ distribution,  management or usage, i.e. smart grid technologies including hybrid vehicles \\ interoperability. \end{tabular} \\ \bottomrule
\end{tabular}
}
\end{table}

The contributions to eco-innovation presented in this review rely mainly on 
the Y02/Y04S tagging scheme of the CPC and the OECD \textsc{Env-Tech} classification, based on a mixture of IPC and CPC codes, which cover a wide range of a wide range technologies related to sustainability objectives. These include energy efficiency in buildings, energy generation from renewable sources, sustainable mobility, and smart grids.

In response to the increasing attention and concerns about climate change mitigation and renewable energy generation, there has been a large increase in the number and scope of patent applications in environment-related domains in the recent past \citep{epo2013sustainable}.

However, searching for environment-related patent documents was not straightforward at the beginning because a dedicated classification system for sustainable technologies was not available. In fact, before 2011, no specific branch of the IPC or of other technology classification covered environment-related inventions. A permanent solution to this issue became available in 2013 following the publication of the Cooperative Patent Classification, which was jointly developed by the the EPO and the United States Patent and Trademark Office (USPTO), in an attempt to harmonize their patent classification procedures. Since then, the CPC has become increasingly popular as a classification standard, complementing or substituting the IPC in a growing number of patent offices worldwide.

As illustrated in Table~\ref{tab:tech_sections}, two types of codes can be found in the CPC classification: codes starting with the letters A to H, which are similar to IPC codes and represent the traditional classification of technologies; and codes starting with Y, which are used to tag cross-sectional technologies already indexed somewhere else in the classification
\footnote{USPTO-CPC Section Y. https://www.uspto.gov/web/patents/classification/cpc/html/cpc-Y.html}. In addition to the Y02 class, the new subclass Y04S dedicated to smart grids was integrated into the CPC section Y. As shown in Table \ref{tab:1}, the Y02 class consists of more than 1000 tags related to sustainable technologies organized in 9 sub-classes. 

The enduring popularity of the IPC classification and its predominant coverage of patents filed before the publication of the CPC prompted efforts to maximize the informative content on eco-innovation present both in the IPC and CPC classifications over a longer time period. In this context, in 2015 the OECD \citep{hascic2015measuring} developed \textsc{Env-Tech}, an expert-based catalogue of environment-related technologies based on the IPC classification, last updated in 2016 \citep{envtech2016}, which can be used to tag green patent documents in  PATSTAT or other patent datasets.
\textsc{Env-Tech} identifies 94 environment-related technology areas that group 4- to 16-digit IPC and CPC codes, building on the CPC Y02 class whenever possible. The catalogue relies on a keyword search strategy identifying patent documents that correspond to each 'target environmental technology field' \citep[p.19]{hascic2015measuring}, covering CCMTs but also environmental management and water-related adaptation technologies (class 1 and 2). When it is not possible to identify single IPC/CPC classes that define alone the technological field of interest, it employs a combination of different IPC/CPC patent classes, also at different aggregation levels. 

Finally, in addition to the CPC and ENV-TECH classifications, the World Intellectual Property Organization's IPC Committee of Experts proposed the \textit{IPC Green Inventory} (IPC-GI).h \footnote{\url{https://www.wipo.int/classifications/ipc/green-inventory/home}, accessed February 2024.} IPC-GI streamlines patent searches for Environmentally Sound Technologies, as identified by the United Nations Framework Convention on Climate Change, with codes spread across various IPC technical fields and the purpose of consolidating them into a single, easily accessible repository. However, given its limited coverage, the IPC calls for caution in its use for research.

%% file: tab_ec_environment_sht.tex
\scriptsize

\begin{longtable}{@{}p{1.7cm}p{3cm}p{3cm}p{2cm}p{4cm}p{2cm}p{4cm}@{}}
\caption{Literature on the relationship between EC and environmental variables}
\label{tab:ec_environment}\\
\toprule
\multicolumn{1}{c}{\textbf{Article}} &
  \multicolumn{1}{c}{\textbf{Dep. var.}} &
  \multicolumn{1}{c}{\textbf{Indep. var.}} &
  \multicolumn{1}{c}{\textbf{Geo}} &
  \multicolumn{1}{c}{\textbf{Rel. EC $\rightarrow$ Env.}} \\* \midrule
\endfirsthead
\multicolumn{7}{c}%
{{\bfseries Table \thetable\ continued from previous page}} \\
\toprule
\multicolumn{1}{c}{\textbf{Article}} &
  \multicolumn{1}{c}{\textbf{Dep. var.}} &
  \multicolumn{1}{c}{\textbf{Indep. var.}} &
  \multicolumn{1}{c}{\textbf{Geo}} &
  \multicolumn{1}{c}{\textbf{Rel. EC $\rightarrow$ Env.}} \\* \midrule
\endhead
\citet{Neagu2019} &
  CO$_2$ emissions &
  ECI, Energy intensity &
  25 EU countries &
  Non-linear: ECI-based EKC \\* \midrule
\citet{Dong2020} &
  Coal consumption &
  Industrial Complexity Index (ICI) and Relatedness density (ECI-based); Energy-saving policies &
  Chinese provinces &
  Linear: $\uparrow$ ICI, $\downarrow$ Coal cons.; energy-saving policies associated to pollution haven hypothesis \\* \midrule
\citet{Can2017} &
  CO$_2$ emissions &
  Exogenous ECI, Energy consumption &
  France &
  Linear: $\uparrow$ ECI, $\downarrow$ CO$_2$; GDP-based EKC (controlling for ECI) \\* \midrule
\citet{Romero2021} &
  GHG intensity, GHG per capita, Product Emission Intensity Index &
  ECI &
  67 countries &
  Linear: $\uparrow$ ECI, $\downarrow$ GHG $\downarrow$ GHGpc $\downarrow$ PEII; centre of the product space has lower emissions \\* \midrule
\citet{Akadiri2022} &
  Ecological footprint (composite index) &
  ECI, Renewable and non-renewable energy use, Economic growth &
  China &
  Linear and two-way: $\uparrow$ ECI, $\uparrow$ Footprint $\uparrow$ Energy cons. \\* \midrule
\citet{Pata2021} &
  Ecological footprint (composite index) &
  ECI, Economic growth, Energy consumption, Globalisation &
  US &
  Non-linear: ECI-based EKC \\* \midrule
\citet{Dordmond2021} &
  Green Jobs Index (GJI) &
  ECI &
  Brasilian states &
  Linear: $\uparrow$ ECI, $\uparrow$ GJI \\* \midrule
\citet{Zheng2021} &
  CO$_2$ emissions &
  ECI, Economic growth, Renewable energy consumption &
  16 leading exporters &
  Linear and two-way: $\uparrow$ ECI, $\downarrow$ CO$_2$; GDP-based EKC \\* \midrule
\citet{Martins2021} &
  CO$_2$ emissions &
  ECI, Economic growth, Energy consumption, Globalisation &
  7 top ECI countries &
  Linear: $\uparrow$ EC, $\uparrow$ CO$_2$; GDP-based EKC \\* \midrule
\citet{Rafique2022} &
  Ecological footprint (composite index) &
  ECI, Human capital, Renewable energy, Urbanization, Economic growth, Export quality, Trade &
  Top 10 ECI countries &
  Linear: $\uparrow$ ECI, $\uparrow$ Footprint \\* \midrule
\citet{Boleti2021} &
  Environmental Performance Index (EPI) &
  ECI/Fitness &
  88 low and high income countries &
  Linear: $\uparrow$ ECI/Fitness, $\uparrow$ EPII; GDP-based EKC  \\* \midrule
\citet{Dogan2021} &
  CO$_2$ emissions &
  ECI, Renewable energy consumption, Non-renewable energy consumption &
  OECD Countries &
  Linear: $\uparrow$ ECI, $\downarrow$ CO$_2$ \\* \midrule
\citet{You2022} &
  CO$_2$ emissions &
  ECI/Fitness, Economic growth &
  95 low, middle and high income countries &
  Linear: $\uparrow$ ECI/Fitness, $\downarrow$ CO$_2$ (high-income); $\uparrow$ ECI/Fitness, $\uparrow$ CO$_2$ (low- and middle-income) \\* \midrule
\citet{Yilanci2020} &
  Ecological footprint (composite index) &
  ECI, Energy consumption, Economic growth &
  China &
  Linear: $\uparrow$ ECI, $\uparrow$ Footprint; ECI-based EKC rejected \\* \midrule
\citet{Swart2020} &
  Deforestation, Forest fires, Solid waste generation, Air pollution &
  ECI &
  Brazilian metropolitan regions &
  Linear: $\uparrow$ ECI, $\uparrow$ Fires $\downarrow$ Waste; GDP-based EKC  \\* \midrule
\citet{Stojkoski2023} &
  GDP growth, Income inequality, CO$_2$ intensity &
  ECI/Fitness &
  World countries &
  Predicting CO$_2$ intensity: Multidimensional EC (trade + technologies + publications) outperforms trade-based; Fitness outperforms ECI \\* \bottomrule
\end{longtable}

\normalsize

%% file: tab_green_prod.tex
\scriptsize

\begin{longtable}{@{}p{1.7cm}p{2.cm}p{2.cm}p{3.5cm}p{3.5cm}p{2cm}p{4cm}@{}}
\caption{Literature for EC based green products classifications}
\label{tab:green_prod}\\
\toprule
\multicolumn{1}{c}{\textbf{Article}} &
  \multicolumn{1}{c}{\textbf{EC indicator}} &
  \multicolumn{1}{c}{\textbf{EC methods}} &
  \multicolumn{1}{c}{\textbf{Geo}} &
  \multicolumn{1}{c}{\textbf{Main Results}} \\* \midrule
\endfirsthead
\multicolumn{7}{c}%
{{\bfseries Table \thetable\ continued from previous page}} \\
\toprule
\multicolumn{1}{c}{\textbf{Article}} &
  \multicolumn{1}{c}{\textbf{Indicator}} &
  \multicolumn{1}{c}{\textbf{Methods}} &
  \multicolumn{1}{c}{\textbf{Geo}} &
  \multicolumn{1}{c}{\textbf{Main Results}} \\* \midrule
\endhead
\citet{fankhauser2013will} &
  Green Innovation Index &
  Descriptive statistics &
  China, Germany, UK, USA, France, Italy, Japan, South Korea &
  Characterization of areas of green potential/weakness  \\* \midrule
\citet{hamwey2013mapping} &
  Products proximity &
  Product space &
  Brazil &
  Full description of Brazilian product space and opportunities for new green production \\* \midrule
\citet{fraccascia2018green} &
  Green product proximity &
  Product space and linear regression &
  141 countries &
  Products proximity is significant for new new green product export at 4-year horizon \\* \midrule
\citet{perez2021mapping} &
  Green Complexity Index and Potential &
  Product space &
  Regions of Mexico &
  Full characterization of Mexican green potential. Confirmed positive correlation between GCI and GCP. Possible scale-variant effects. \\* \midrule
\citet{mealy2022economic} &
  ECI, Green Complexity Index and Potential &
  Product space &
  141 countries &
  Green products are more complex, negative correlation between GCI and CO$_2$/cap emission \\* \midrule
\citet{muller2023economic}&
  PCI, GCI &
  Product space &
  141 countries, special focus on EU and MENA regions &
  Extended list of product relevant for green hydrogen production via proximity in the product space. Measure of GCI for green hydrogen-related product of EU and MENA countries\\*
 \bottomrule
\end{longtable}

\normalsize

%% file: tab_green_tech.tex
\scriptsize
\setlength\LTleft{-4.7cm}
\begin{longtable}
{@{}p{2cm}p{2.2cm}p{2.4cm}p{3cm}p{1.8cm}p{3cm}p{1.3cm}p{3.5cm}p{1.9cm}@{}}
\caption{Literature on green technologies}
\label{tab:green_tech}\\
\toprule
  \multicolumn{1}{c}{\textbf{Article}} &
  \multicolumn{1}{c}{\textbf{Subtopic}} &
  \multicolumn{1}{c}{\textbf{Dep. var.}} &
  \multicolumn{1}{c}{\textbf{Indep. var.}} &
  \multicolumn{1}{c}{\textbf{EC def.}} &
  \multicolumn{1}{c}{\textbf{\begin{tabular}[c]{@{}l@{}}Relatedness \\ def.\end{tabular}}} &
  \multicolumn{1}{c}{\textbf{Geo}} &
  \multicolumn{1}{c}{\textbf{Main Results}} &
 \textbf{\begin{tabular}[c]{@{}l@{}}Dataset;\\ Geoloc.;\\ Classification\end{tabular}}\\ \midrule

\endfirsthead
\citet{montresor2020green} &
Green and non-green  technology and key  enabling technologies  (KETs) &
New RTAs in green technologies  &
Relatedness of green technology to all, non-green, green  technologies; new RTAs in KETs, interaction between new RTAs in KETs and relatedness  & &
Proximity-based relatedness of tech. $z$ at time $t$ in region $i$ to all technologies at $t$;  green or non-green  technologies at $t - 5$&
EU NUTS2  regions &
Relatedness to non-green (rather than green) knowledge makes new  green-tech specialisation more likely; KETs favour transition to green  technologies and moderate relatedness to pre-existing technologies' impact &
REGAPT (applicants);
REGPAT  NUTS3;
ENV-TECH  \\ \midrule

\citet{barbieri2020specialization} &
  Green technology  life cycle, related  and unrelated variety &
  Green technology patent families (all and at different life cycle stages)&
  Green technology related, semirelated,  unrelated variety &
  Technology  ubiquity &
  Related, semirelated, unrelated variety of green technologies computed with Shannon entropy if they share 
  IPC code at different aggregation levels &
  US states &
  Unrelated variety positive predictor of green-tech specialisation;   
  unrelated variety main driver of green technology development over the life cycle&
  PATSTAT  2016 (applicants);
  GeoNames, Google Maps API; ENV-TECH   \\ \midrule

\citet{perruchas2020specialisation} &
  Green technological  specialisation and  life cycle across  countries &
  New RTAs in green technologies  (all/only if country  has RTA \textgreater{}1 at t-1) &
  Relatedness,  technology life cycle,  environmental  policy stringency &
  ECI (green technology complexity)&
  Proximity-based relatedness of green tech. $z$ in which country $i$ specialised  at time $t$ to technologies in 
 which $i$ already specialised at $t$ &
  63 countries &
 Diversification  in green technologies  related to pre-existing  competences and more associated to green  technology maturity than income, with green techn complexity  not hampering  further specialisation &
  PATSTAT  2016 (applicants);
  GeoNames,   Google Maps API;
  ENV-TECH \\ \midrule

\citet{santoalha2021digital} &
Green technology and digital skills 
  &
New RTAs in green  or non-green technologies &
Digital literacy, overall, green to non-green relatedness & &

Proximity-based relatedness of green tech. $z$ in which region $i$ not specialised  at time $t$ to technologies in which  $i$  already specialised at $t$ 
  &
EU NUTS2  regions 
  &
Digital skills have a positive impact on a region's ability to specialize in new technologies, especially in green domains, with e-skills moderating the effect of relatedness
  &
  REGPAT (applicants);
  REGPAT  NUTS3;
  ENV-TECH \\ \midrule

\citet{santoalha2021diversifying} &
Green technology and national or regional environmental policy support 
  &
New RTAs in green technologies
  &
National and regional environmental policy support, green to non-green relatedness 
  &
  &
Proximity-based relatedness of green tech. $z$ in which region $i$ not specialised at time $t$ to technologies in which $i$ specialised at $t$ 
  &
EU NUTS2  regions 
  & 
Related technological capabilities more important than regional environmental policies, while national political support for environmental policies mitigates the importance of related technologies
  &
REGPAT (applicants);
REGPAT  NUTS3;
ENV-TECH \\ \midrule
  
  \citet{bergamini2020exploring} &
Potential for green technology development and socio-economic regional characteristics
  &
RTAs in green technologies at time $t$ 
  &  
Potential technological advantage \citep{hausmann2022implied}; labour
market participation, STEM employment, education, R\&D  expenditures etc. 
  &
  &
Proximity-based relatedness of green tech. $z$ in which region $i$ not specialised  at time $t$ to all technologies in which $i$ already specialised at  $t$ 
  &
EU NUTS2  regions 
  & 
Green techns RTAs concentrated in FR, DE, Northern IT, increase over time, but depend on technology type. Future regional green RTAs positively associated with labour market participation rate, employment duration, STEM employment, R\&D exp., higher education
  &
REGPAT (inventors and applicants);
REGPAT  NUTS3;
 \citet{fiorini2017monitoring} classif.
based on Y02/Y04S\\ \midrule

\citet{sbardella2018green} &
Green innovative  capacity across  countries & & &
EFC  (Green  Technology  Fitness) & 
  &
72 countries 
  & 
Global set of green technological  capabilities is sticky with stable leaders, catch-ups  (especially in East Asia) and laggards. Direct relationship  between income and green  innovative capacity, with  complexity resonating  with green technology   life cycle &
PATSTAT  2016 (applicants);  GeoNames,  Google Maps  API;
ENV-TECH \\ \midrule

\citet{napolitano2022green}&
  Green innovative  capacity and income inequality &
  Green Technology  Fitness & Income inequality& 
  EFC  (Green  Technology  Fitness) & &
  57 countries &
  Green innovative capacity negatively correlated to  income inequality, but  inequality irrelevant for high-GDP countries,  while for middle-income  countries lower inequality  favours green specialisation &
  PATSTAT  2016 (applicants);  GeoNames,  Google Maps  API;
  ENV-TECH \\ \midrule

\citet{de2022trickle}&
  Green technology and exported goods &
  RCAs  in exported products &
  RTAs in green  technologies & 
  EFC  (Green  Technology  Complexity) &

Multi-layer based relatedness of green technology $z$ at time $t$ or $t-5$ to product $p$ at $t$ &

48 countries &

Strong connection between  green technologies and  exported products related to the  processing of raw materials;  with time, more complex green  knowledge trickles down into  more complex goods &
REGPAT (applicants); REGPAT countries; Y02/Y04S  
    \\ \midrule

\citet{barbieri2022regional} &
  Green and non-green  know-how in EU regions &
  Green Technology  Fitness &
  Non-green Technology  Fitness,  Non-green to  
  green relatedness &
  EFC  (Green and  Non-green Technology  Fitness) &
  Multi-layer based relatedness of non-green tech. $z$ in which 
  region $i$  specialised at time $t-5$  to green tech.  $z$ in which $i$ already specialised at $t$ &
  EU NUTS2  regions &
Non-green and green competitiveness geographically heterogeneous and stable over time (persistent between Central-Eastern EU dichotomy). Complementarity between non-green and green tech know-how, but green specialisation depends on regional patent portfolios' composition &
  PATSTAT  (inventors);
  PATSTAT  inventor  residence,  \cite{de2019geocoding}'s geocoding,  EUROSTAT  GIS NUTS 3; Y02/Y04S    \\ \midrule

  \citet{grashof2023dark} &
  Regional divergence, relatedness between brown and green know-how in EU regions &
  New RTAs in green technologies (from time $t-5$ to $t$)  &
  Relatedness to green or brown technologies (at time $t$), economic strength ($=1$ if region $i$'s GDP per capita $\geq$ 75\% of EU  average, $=0$ otherwise)&
  ECI  (brown and green Technology Complexity) &
  Proximity-based relatedness of brown/green technology $z$ in which region $i$  not specialised at time $t$ to technologies in which $i$ specialised at $t$ &
  EU NUTS2  regions&

Related technological know-how more important for green specialisation than economic strength; high-income regions higher green relatedness, but low-income regions with high green relatedness can successfully diversify into complex green tech  &
  REGPAT (inventors);
  REGPAT NUTS2; WIPO IPC Green/Brown Inventory\\ 
  \bottomrule
\end{longtable}
\normalsize

%% file: appendix-patent_data.tex
In the following, we provide additional information on patent data and their potential use for analysing climate change mitigation and adaptation technologies. 

As mentioned in the paper, the two main data sources on countries' patenting activity are PATSTAT and REGPAT, on which we offer additional details in the following, especially aimed at a comparison between these two data sources. 

\begin{figure}[h]
    \centering
    \includegraphics[width=0.95\textwidth]{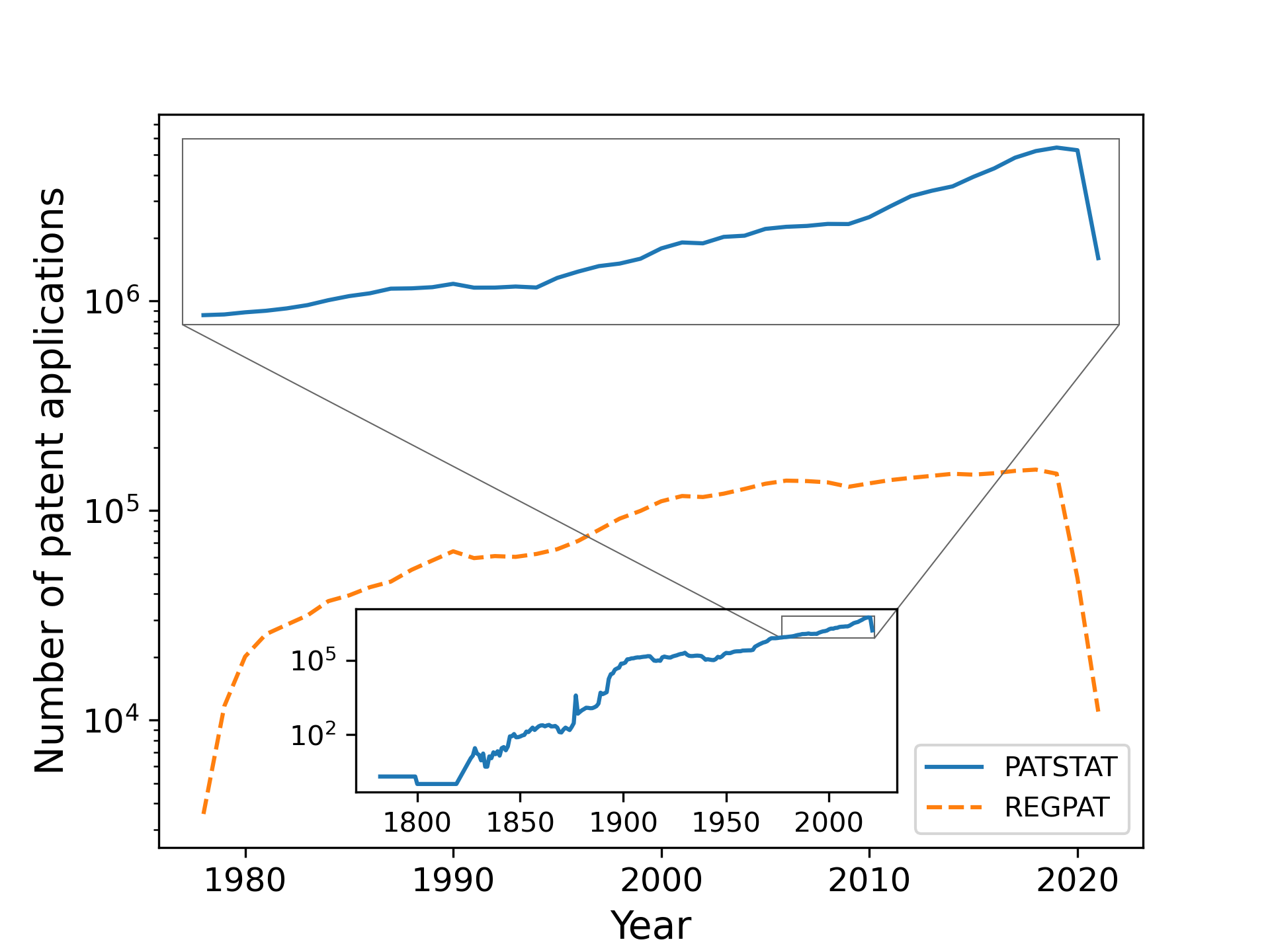}
    \caption{Comparison of the coverage of PATSTAT (blue) and REGPAT (orange). Main plot: time series of the number of patent applications recorded in PATSTAT and REGPAT over the period covered by both databases (1978-2022). Inset: time series of the number of patent applications recorded in PATSTAT over the entire period it covers.}
    \label{fig:PATSTAT_vs_REGPAT}
\end{figure}

In fact, as can be appreciated in Figure \ref{fig:PATSTAT_vs_REGPAT}, PATSTAT and REGPAT 
present significant differences in their temporal and geographical coverage. 
The large difference in the number of documents is due to the fact that REGPAT records patents filed only at the EPO, while PATSTAT collects information from most patent offices worldwide. This makes REGPAT less suitable for the analysis of, for instance, smaller countries. However, the EPO tends to receive high-quality applications, making data collected from it more reliable. 

While REGPAT is published annually by the OECD and covers the subset of PATSTAT patent applications filed only at the EPO since its inception in 1978, PATSTAT has been published biannually by the European Patent Office (EPO) since 2007 and has grown substantially in coverage over time. As of the latest editions, it records information on more than 100 million patent applications filed since the late eighteenth century, which are collected in over 50 million families.
Whenever the information is available for a patent application, PATSTAT records, among other things, the receiving patent office, the filing date, the technologies in which the patent innovates (encoded in standard technology codes), and the residence of the applicants and of the inventors at the time of filing.
The geographical information is incomplete, with the coverage varying widely across patent offices. The most commonly available information in this sense is the country of residence of inventors and applicants. 

Since, for geographical reasons, the EPO attracts disproportionately more European patent applications, it does not provide a uniform geographical coverage beyond European borders. Furthermore, filing costs at the EPO are higher than in most national offices. This skews the sample towards ``high-value'' patents and, indirectly, leads to an over-representation of richer European countries.
However, REGPAT makes up for this shortcoming thanks to an accurate geocoding across over 5500 sub-national regions of the patent documents filed by applicants or inventors residing in one of over 40 countries belonging to the OECD, the EU, the UK, Brazil, China, India, the Russian Federation, and South Africa.\footnote{The brochure of an older version of REGPAT, available at \url{http://econ.geo.uu.nl/crespo/tech_div/regpat_201602.pdf}, provides more details on the structure of the geographical coverage}

Concerning the temporal coverage, Figure \ref{fig:PATSTAT_vs_REGPAT} shows that REGPAT and PATSTAT differ substantially. However, it also suggests that both databases cover a long enough time window for all practical purposes. The coverage of REGPAT starts only in 1978, while PATSTAT records US and UK patents dating back as far as the late eighteenth century. Nevertheless, large numbers of patents have been recorded only in recent decades. Moreover, both databases are updated to virtually the same date. 

It is worth noting that, due to the dynamics of patent offices and the regulations governing patent filing, there is an intrinsic lag of 12-18 months between when an application is received by a patent office and when the corresponding record appears in the database, at the end of the so-called search phase by the patent office.
During the search phase, patent offices conduct a primary assessment of the originality of the invention and inform the applicants, who can decide to either withdraw the application thus keeping it confidential, or pursue the grant of a patent and disclose their application to the public. 
Therefore, patent counts extracted from an edition of PATSTAT or REGPAT published in 2022 are not reliable beyond 2018 or 2019.

An invention can be submitted to different offices –- e.g. to cover different geographical regions -- by filing different applications at different points in time.
The first patent application filed to protect an invention is called the priority application.
Subsequent applications that are related to the same invention name the same applications as priorities, allowing to group into the same patent family (there are over 50 million patent families in PATSTAT).
Families are a useful way to group together documents referring to the same innovation and are thus frequently used as the basic unit of observation in empirical exercises.

As mentioned above, PATSTAT provides limited information about the geographical location of inventors and applicants for many documents, even though the database records patents filed by applicants and inventors located in more than 200 countries.
Instead, REGPAT offers much more detailed information, albeit on a smaller set of countries, by associating patent documents to the OECD Territorial Level (TL) \citep{oecd-geo} code of the region of residence of applicants and inventors.
For European countries the TL follows the hierarchical structure of the 2013 edition of the Nomenclature of Territorial Units for Statistics\footnote{\url{https://ec.europa.eu/eurostat/web/gisco/geodata/reference-data/administrative-units-statistical-units/nuts}}) (NUTS), developed by Eurostat. Instead, in the US, progressively finer levels of the classification identify states, economic areas and counties, that do not follow a nested structure.

To attribute a geographical location to patents one can leverage the inventors' as well as the applicants' residence information. The former is often preferred as a proxy for the location of inventive capabilities because it is assumed that inventors (who are always physical persons) tend to live close to where they perform their duties. On the contrary, applicants (who are in many cases companies) may choose to assign to the corporate headquarters a patent that was developed in a subsidiary located in a different country or region for business-related reasons.

Multiple inventors or applicants can be linked to the same patent application or family. In such cases, one may choose between counting the patent fractionally and counting patents in full, when it comes to assigning patents to geographical areas. Full counting counts as a unit of every pair of patent documents and territorial units hosting at least one inventor. As a consequence of this double counting, the weight attributed to a patent (and even more for a family) depends on the number of inventors.
Instead, in fractional counting, each patent (or each family) sums to 1, and each territorial unit having an inventor gets a fractional weight inversely proportional to the total number of inventors.
There is some debate in the literature concerning the best approach~\citep{Waltman2016}. However, in economic complexity applications, fractional counting is generally the preferred approach.

\subsection{Identification of green technologies}

However, searching for environment-related patent documents was not straightforward at the beginning because
a dedicated classification system for sustainable technologies was not available. 
In fact, before 2011, no specific branch of the IPC or of other technology classification systems
covered environment-related inventions.

A first step in this direction was the creation in 2011 of the Y02
class the EPO, in cooperation with the United Nations Environmental Program (UNEP) and the International Centre on Trade and Sustainable Development (ICTSD), to complement the IPC classification System. 
From the beginning, the purpose of the Y02 class was to tag CCMT patent documents by means of search strategies and algorithms implemented by expert examiners and that can be re-run periodically to update the classes \citep{veefkind2012new}. The Y02 scheme initially covered only patent documents related to CCMTs in the energy sector and was later extended also to other types of mitigation technologies. 
This effort by the EPO constituted  a major advancement for the study of green innovation both from an academic and policy perspective, as it has allowed also non-specialists to easily identify CCMTs. 

In 2013, the European Patent Office and the United States Patent and Trademark Office (USPTO) agreed to harmonize their patent classification practices and developed the Cooperative Patent Classification (CPC). Since then, the CPC has become increasingly popular as a classification standard and has been complementing or substituting the IPC in a growing number of patent offices worldwide.
As illustrated in Table~\ref{tab:tech_sections}, two types of codes can be found in the CPC classification: codes starting with the letters A to H -- similar to IPC codes and representing the traditional classification of technologies; and codes starting with Y, which are used to tag cross-sectional technologies that are already indexed somewhere else in the classification
\footnote{USPTO-CPC Section Y. https://www.uspto.gov/web/patents/classification/cpc/html/cpc-Y.html} Therefore, the Y classification scheme is used to tag patent documents that are already classified or indexed somewhere else in the classification. 
In addition to the Y02 class, the new subclass Y04S dedicated to smart grids
was integrated in the CPC section Y.
As shown in Table \ref{tab:1}, the Y02 class consists of more than 1000 tags related to sustainable technologies organized in 9 sub-classes. 

With the aim of maximizing the informative content on eco-innovation present both in the IPC and CPC classifications, in 2015 the OECD \citep{hascic2015measuring} developed \textsc{Env-Tech}, an expert-based catalogue of environment-related technologies based on the IPC classification and lastly updated in 2016 \citep{envtech2016}, which can be used to tag green patent documents in  PATSTAT or other patent data-sets.
\textsc{Env-Tech} identifies 94 environment-related technology areas that group 4- to 16-digit IPC and CPC codes, building on the CPC Y02 class whenever possible. The catalogue relies on a keyword search strategy identifying patent documents that correspond to each "target environmental technology field'' \citep[p.19]{hascic2015measuring} covering CCMTs but also environmental management and water-related adaptation technologies (class 1 and 2). When it is not possible to identify single IPC/CPC classes that portray alone the technological field of interest, it employs a combination of different IPC/CPC patent classes, also at different aggregation levels. 

While widening the object of analysis is undoubtedly commendable, unfortunately, it makes the catalogue very vulnerable to patent document re-classification. In fact, while the additional information may have been useful for the researchers working with versions of PATSTAT prior or contemporaneous to 2016, the year up to which \textsc{Env-Tech} is updated, using a combination of different IPC and CPC codes may constitute a drawback when working with more recent versions of PATSTAT. This is because the classification of the codes comprising any technology -- especially for finer-grained codes -- changes over time, possibly reclassifying past inventions into new codes. Furthermore, being fixed in time, \textsc{Env-Tech} not only it suffers from the reclassification of previous inventions, but it also fails to consider newer patent applications. In fact, technology codes are revised at least once a year to take into account new technical advances, as well as to improve the search for the prior art that the patent officers use to establish the application's degree of innovation. 
Moreover, the Y04S subclass covering smart grids, which at the end of 2016  comprised 54000 patent documents \citep{angelucci2018supporting}, is not included in the \textsc{Env-Tech} catalogue. By contrast, the Y02/Y04S codes are part of the CPC section Y and are therefore robust to changes in the CPC classification, are more user-friendly and readily usable for the relevant data can be directly extracted by PATSTAT with no need for intermediate steps. 
Lastly, another limitation of \textsc{Env-Tech} is that, due to its limited granularity, it does not allow very detailed studies, an obstacle for carrying out economic complexity analyses that usually rely on very fine-grained information. 